\documentclass[fleqn,usenatbib]{mnras}

\usepackage{newtxtext,newtxmath}

\usepackage[T1]{fontenc}
\usepackage{ae,aecompl}

\usepackage{graphicx}	
\usepackage{amsmath}	
\usepackage{subcaption}
\usepackage{pdflscape}
\usepackage{supertabular}

\usepackage{eso-pic}

\AddToShipoutPictureBG*{%
  \AtPageUpperLeft{%
    \hspace{0.82\paperwidth}%
    \raisebox{-5\baselineskip}{%
      \makebox[0pt][l]{\textnormal{DES-2023-0751}}
}}}%

\AddToShipoutPictureBG*{%
  \AtPageUpperLeft{%
    \hspace{0.73\paperwidth}%
    \raisebox{-6\baselineskip}{%
      \makebox[0pt][l]{\textnormal{FERMILAB-PUB-23-381-PPD}}
}}}%

\title[OzDES RM stacking analysis]{OzDES Reverberation Mapping Program: Stacking analysis with H$\beta$, Mg\,\textsc{ii} and C\,\textsc{iv} }

\author[Malik et al.]{
\parbox{\textwidth}{
\Large
U.~Malik$^{1}$\thanks{umang.malik@anu.edu.au},
R.~Sharp$^{1}$\thanks{rob.sharp@anu.edu.au},
A.~Penton$^{2}$,
Z.~Yu$^{3}$,
P.~Martini$^{3,4}$,
B.~E.~Tucker$^{1,5,6}$,
T.~M.~Davis$^{2}$,
G.~F.~Lewis$^{7}$,
C.~Lidman$^{1,8}$,
M.~Aguena$^{9}$,
O.~Alves$^{10}$,
J.~Annis$^{11}$,
J.~Asorey$^{12}$,
D.~Bacon$^{13}$,
D.~Brooks$^{14}$,
A.~Carnero~Rosell$^{9,15,16}$,
J.~Carretero$^{17}$,
T.-Y.~Cheng$^{18}$,
L.~N.~da Costa$^{9}$,
M.~E.~S.~Pereira$^{19}$,
J.~De~Vicente$^{20}$,
P.~Doel$^{14}$,
I.~Ferrero$^{21}$,
J.~Frieman$^{11,22}$,
G.~Giannini$^{17}$,
D.~Gruen$^{23}$,
R.~A.~Gruendl$^{24,25}$,
S.~R.~Hinton$^{2}$,
D.~L.~Hollowood$^{26}$,
D.~J.~James$^{27}$,
K.~Kuehn$^{28,29}$,
J.~L.~Marshall$^{30}$,
J. Mena-Fern{\'a}ndez$^{20}$,
F.~Menanteau$^{24,25}$,
R.~Miquel$^{17,31}$,
R.~L.~C.~Ogando$^{32}$,
A.~Palmese$^{33}$,
A.~Pieres$^{9,32}$,
A.~A.~Plazas~Malag\'on$^{34,35}$,
K.~Reil$^{35}$,
A.~K.~Romer$^{36}$,
E.~Sanchez$^{20}$,
M.~Schubnell$^{10}$,
M.~Smith$^{37}$,
E.~Suchyta$^{38}$,
M.~E.~C.~Swanson$^{14}$,
G.~Tarle$^{10}$,
C.~To$^{4}$,
N.~Weaverdyck$^{10,39}$,
P.~Wiseman$^{37}$
\\
{\it \footnotesize (Affiliations listed at the end of the paper)}
}
}

\date{Accepted 2024 April 25. Received 2024 April 21; in original form 2023 July 31}

\pubyear{2024}

\begin{document}
\label{firstpage}
\pagerange{\pageref{firstpage}--\pageref{lastpage}}
\maketitle

\begin{abstract}
Reverberation mapping is the leading technique used to measure direct black hole masses outside of the local Universe. Additionally, reverberation measurements calibrate secondary mass-scaling relations used to estimate single-epoch virial black hole masses. The Australian Dark Energy Survey (OzDES) conducted one of the first multi-object reverberation mapping surveys, monitoring 735 AGN up to $z\sim4$, over 6 years. The limited temporal coverage of the OzDES data has hindered recovery of individual measurements for some classes of sources, particularly those with shorter reverberation lags or lags that fall within campaign season gaps. To alleviate this limitation, we perform a stacking analysis of the cross-correlation functions of sources with similar intrinsic properties to recover average composite reverberation lags. This analysis leads to the recovery of average lags in each redshift-luminosity bin across our sample. We present the average lags recovered for the H$\beta$, Mg\,\textsc{ii} and C\,\textsc{iv} samples, as well as multi-line measurements for redshift bins where two lines are accessible. The stacking analysis is consistent with the Radius-Luminosity relations for each line. Our results for the H$\beta$ sample demonstrate that stacking has the potential to improve upon constraints on the $R-L$ relation, which have been derived only from individual source measurements until now.  
\end{abstract}




\section{Introduction}

Reverberation mapping is a powerful technique that can resolve the cores of active galactic nuclei (AGN) in the time domain. The accretion disk around the central supermassive black hole (SMBH) produces intrinsically variable emission at UV-optical wavelengths. The surrounding broad-line region (BLR) is ionised by this continuum emission, which drives a reverberation response in the emission-line flux from the BLR on time-scales of weeks to months \citep{Blandford1982,Peterson1993}. Multi-epoch photometric and spectroscopic observations can be used to trace the variability of the continuum and the emission-line response, respectively. 

The size difference between the accretion disk (light-days to weeks) and the BLR (light-weeks to months) introduces a delay in the response of the BLR to the variation in the ionising flux. The delay, \textit{i.e.} reverberation lag, $\tau$, can be recovered by cross-correlating the two light curves, in order to measure the radius of the BLR ($R_{\rm BLR} = c\tau$). The velocity dispersion of the BLR ($\Delta V$) can be estimated from the width of the broadened emission lines. The mass of the central black hole ($M_{\textrm{BH}}$) can then be measured using the virial theorem:
\begin{equation}\label{virial}
    M_{\textrm{BH}} = f\frac{R_{\textrm{BLR}} \Delta V^2}{G},
\end{equation}
where $f$ is the virial coefficient; a dimensionless scale factor that accounts for the geometry, orientation, and kinematics of the BLR \citep{Woo2015}. 

Reverberation mapping (RM) is presently the only method that can be used to directly measure SMBH masses beyond the local Universe, as other techniques are reliant on resolving the gravitational sphere-of-influence of the black hole, which remains challenging even with high angular resolution instruments  \citep[e.g.,][]{Gebhardt2000,Gebhardt2011,Kuo2011,EHT2019}. However, RM is by nature observationally intensive. It requires repeated observation over the relevant variability time-scales of AGN in order to ensure the light-curve variability and reverberation lag are resolved \citep{Horne2004}. Early surveys monitored AGN on a source-by-source basis, making observations over several months to years. Lag measurements were made for dozens of sources, using the H$\beta$ line \citep[e.g.,][]{Peterson1998,Kaspi2000,Peterson2004,Bentz2009}. From these measurements, a tight correlation was found between the AGN luminosity and the radius of the BLR \citep[$R-L$ relation; e.g.,][]{Bentz2009,Bentz2013}. Lags recovered using higher ionisation emission lines (e.g., C\,\textsc{iv}) were found to be shorter than lags recovered using lower ionisation emission lines (e.g.,\,H$\beta$), demonstrating the ionisation stratification of the BLR \citep{Gaskell1986,Dietrich1993}. The $R-L$ relation is importantly used to calibrate secondary mass-scaling relations to estimate single-epoch virial BH masses \citep[e.g.,][]{Shen2011}, and has also been proposed as a way to standardise AGN for use as a cosmological distance indicators \citep{Watson2011,MartinezAldama2019}. 

Through the advent of wide-field photometric surveys such as the Dark Energy Survey, \citep[DES,][]{DES2016}, multi-epoch photometric data for large samples of AGN has become readily available. Concurrent observations can be made with multi-object spectrographs, however the demand for these instruments is high and therefore limits the number of epochs which can be feasibly acquired. The Australian Dark Energy Survey (OzDES) and Sloan Digital Sky Survey Reverberation Mapping (SDSS-RM) Project have conducted the first multi-object RM surveys, observing hundreds of AGN probing a wide range of AGN luminosities and redshifts \citep{King2015,Shen2015}. These programs have delivered over one hundred new lag measurements \citep[][Penton et al. in prep]{Grier2017,Hoormann2019,Grier2019,Homayouni2020,Yu2021,Yu2023,Malik2023}. This allowed the Mg\,\textsc{ii} and C\,\textsc{iv} $R-L$ relations to be constrained for the first time using statistically significant samples, however there is significant scatter in the measurements. This is mostly due to challenges with data quality \citep[low signal-to-noise, limited sampling; see][]{Malik2022} and lag recovery reliability \citep{Li2019,Penton2022}. These factors have limited the lag recovery efficacy of each survey to about 10-25\%. 

Stacking can be used to combine the cross-correlation signals of physically similar AGN to recover average lags for these objects. The technique was first applied by \citet{Fine2012,Fine2013} using only two spectroscopic epochs, which yielded a marginal result. After demonstrating the success of the technique with the \citet{Bentz2013} H$\beta$ sample, \citet{Li2017} measured composite lags with H$\alpha$, H$\beta$, He\,\textsc{ii} and Mg\,\textsc{ii} using a subset of the SDSS-RM sample. Stacked averages are not swayed by the systematic errors from individual sources. Their consistency (or not) with the measurements made for individual sources is therefore important. Additionally, with the wide redshift range covered by our sample, the gaps in the observational window function can be filled in to some extent. Therefore, stacking leverages the data in a way that cannot be done with traditional individual measurements. 

We present a stacked lag analysis of the entire OzDES sample, for the H$\beta$, Mg\,\textsc{ii} and C\,\textsc{iv} lines. Section \ref{sec:data} details the observations obtained by OzDES and the data calibration procedures. In Section \ref{sec:method} we describe the technique used to recover stacked lags. In Section \ref{sec:results} we present our average lag measurements, and comparisons with individual measurements on the respective $R-L$ relationships for each emission line. We summarize our results and the discuss the outlook to the future in Section \ref{sec:summary}.

Throughout this work we adopt a flat $\Lambda$CDM cosmology, with $\Omega_{\Lambda}=0.7$, $\Omega_{M}=0.3$, and $H_0=70$ km\,s$^{-1}$\,Mpc$^{-1}$.

\section{Data} \label{sec:data}

Our photometric data were obtained as part of the Dark Energy Survey (DES) Supernova Program, which observed 10 deep fields covering 27\,deg$^2$, comprising the ELAIS, XMM-LSS, Chandra deep-field South, and SDSS Stripe 82 fields \citep{Kessler2015,Morganson2018}. These fields were observed in the $griz$ filters, with the Dark Energy Camera (DECam) on the 4m Blanco telescope at Cerro Tololo Inter-American Observatory (CTIO) \citep{Flaugher2015}. The fields were observed with $\sim$6 day cadence over a 5-6 month season (August to January) from 2013 to 2018, with additional science verification data taken in 2012. The OzDES project \citep{Yuan2015,Childress2017,Lidman2020} conducted follow-up spectroscopic observations with the 2dF multi-object fibre positioning system and the AAOmega spectrograph \citep[3700-8800\,\AA, ][]{Sharp2006} on the 3.9m Anglo-Australian Telescope (AAT), taken with approximately monthly cadence over the same seasons, from 2013 to 2019. After the conclusion of the Supernova Program, additional DECam observations were taken monthly in the 2018-19 season, taking the baseline of our photometric light curves to 7 years.

\begin{figure}
	\centering
	\includegraphics[width=\columnwidth]{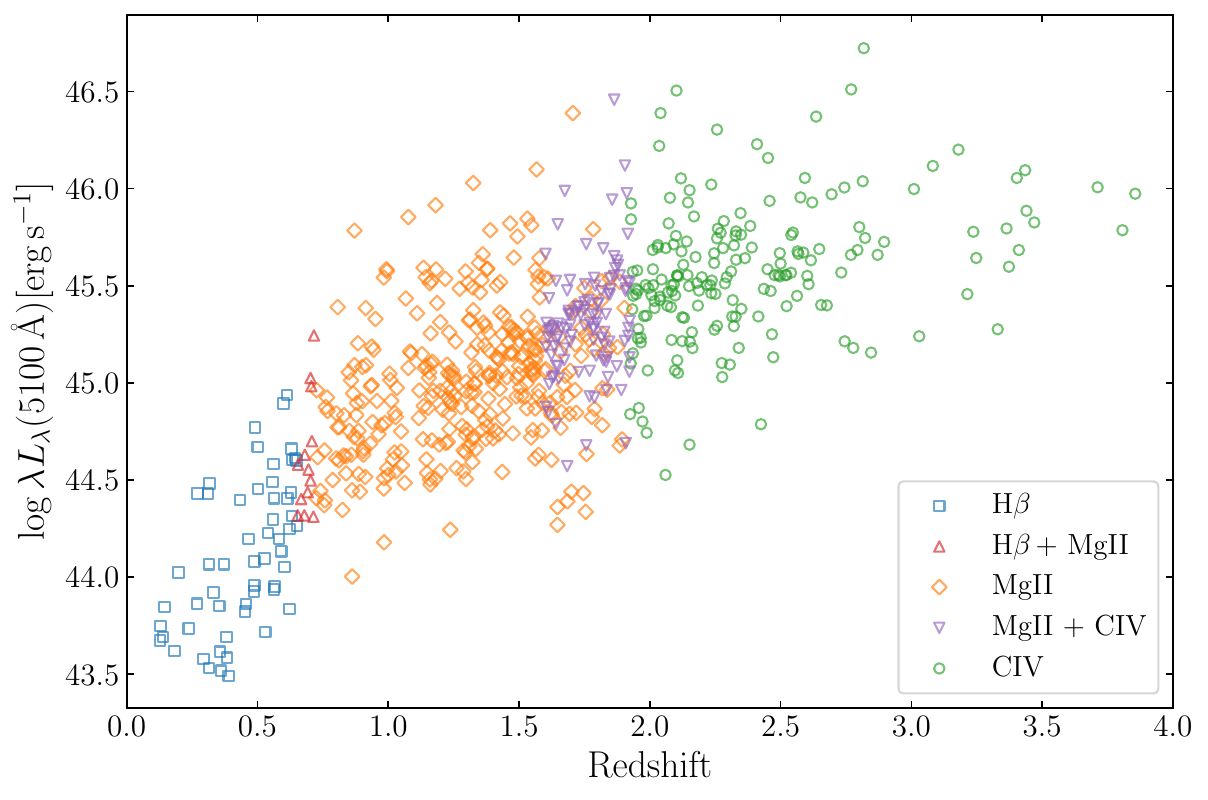}
	\caption{Distribution of redshifts and monochromatic luminosity at 5100\,\AA\ for the 690 AGN in the OzDES RM sample. The H$\beta$ sample extends to $z=0.75$, with 13 sources overlapping with our Mg\,\textsc{ii} sample. The Mg\,\textsc{ii} sample extends to $z=1.92$, with 106 sources overlapping with our C\,\textsc{iv} sample.} 
    \label{fig:ozdes_AGN}
\end{figure}

The OzDES Reverberation Mapping sample comprises 735 AGN, ranging from $0.1<z<4.0$, with apparent magnitudes 17.2 $<$ $r_{\rm{AB}}$ $<$ 22.3 \citep{Tie2017}. Of these 735 AGN, we removed 9 of the 78 AGN from the H$\beta$ sample due to broad-absorption lines (BALs) or incorrect classification as a Type 1 AGN, 3 of the 453 AGN from the Mg\,\textsc{ii} sample for the same reason (these sources overlapped with the H$\beta$ sample), and 88 of the 378 AGN from the C\,\textsc{iv} sample due to BALs. These sources were included in the initial survey selection to improve the source density on the sky and study BAL variability, but have proven challenging for reverberation analysis. The final sample we use in this analysis comprises 69 H$\beta$ sources, 450 Mg\,\textsc{ii} sources, and 290 C\,\textsc{iv} sources (690 AGN in total). The redshift and luminosity distribution of these targets is shown in \autoref{fig:ozdes_AGN}. 

As done for the individual lag measurements made by \citet{Yu2023} and \citet{Malik2023}, for the H$\beta$ and Mg\,\textsc{ii} samples we measure the emission-line flux from spectra obtained on different nights as separate epochs in order to maximise the cadence of our sampling for our emission-line light curves. As we have lower signal-to-noise for the C\,\textsc{iv} sample, we co-add the spectra over each observing run (typically 4-7 nights during dark time each month), as done by \citet{Hoormann2019}. 

We do not measure the continuum luminosity directly from the spectra due to fibre aperture effects from variable atmospheric seeing and fibre placement uncertainties. From the average $r$-band magnitude and redshift of the AGN, we estimated the monochromatic continuum flux at rest frame 5100\,\AA, 3000\,\AA\ and 1350\,\AA\ using the DECam $r$-band filter transmission curve and the SDSS quasar template \citep{VandenBerk2001}. The template is scaled to the magnitude of the source, assuming $L_{\rm bol}=9\, \lambda L_{\lambda}$\,(5100\,\AA) \citep{Kaspi2000}. The source properties for the OzDES sample used in this work are provided in \S\ref{sec:appendixA} and complete sample characteristics for the final OzDES RM sample will be provided in a future OzDES RM Program paper.

The DES photometry is calibrated using the DES data reduction pipeline \citep{Morganson2018,Burke2018}. We perform a spectrophotometric flux calibration following \citet{Hoormann2019}. For the H$\beta$ and C\,\textsc{iv} samples, we measure the line fluxes as done by \citet{Hoormann2019}. For the continuum subtraction, the local continuum windows we use for H$\beta$ are 4760 to 4790\,\AA\ and 5100 to 5130\,\AA, and for C\,\textsc{iv} are 1450 to 1460\,\AA\ and 1780 to 1790\,\AA. For the Mg\,\textsc{ii} sample, the iron subtraction and line flux measurement is performed as detailed in \citet{Yu2023}. The calibration uncertainties of the line flux for each line are measured using the F-star warping function method as detailed in \citet{Yu2021}.

\section{Lag recovery method}\label{sec:method}

As reverberation lags should be dependent on the intrinsic AGN luminosity alone (at least to first order), we bin the sources by their continuum luminosity. Further details are provided for each emission-line subsample in \S\ref{sec:results}. 

For each source in a redshift-luminosity bin, we covert the observation dates to the rest frame of the source by dividing by (1+$z$). We use the \texttt{PyCCF} code to perform the interpolated cross-correlation function method for each individual source \citep[ICCF;][]{Gaskell1987,2018ascl.soft05032S}. The continuum and emission-line light curves of each source are linearly interpolated to a grid spacing of 3 days. The interpolated light curves are cross-correlated as a function of time-lag. We then average the cross-correlation functions (CCF) of each source in the bin to obtain the stacked CCF. We search for lags over a (rest-frame) lag range of $[-100, 300]$ days for H$\beta$ and C\,\textsc{iv}, and $[-100, 500]$ days for Mg\,\textsc{ii} as it has a longer expected lag for our sample. 

We measure the average lag and its uncertainties by bootstrapping the sample in each bin. We perform the above procedure to calculate the stacked CCF's for each bootstrapped re-sample of the original binned sample. We repeat this 1000 times, and record the centroid of each CCF ($r_{\rm max}$) to build the bootstrap distribution, from which we adopt the median and 16th and 84th percentiles of this distribution as the recovered average stacked lag, $\tau$, and lower and upper uncertainties, $\sigma_{\tau}$, for the bin.

Following \citet{Li2017}, we shuffle the spectroscopic epochs and repeat the stacking for 100 Monte Carlo realisations, and compare the stacked CCF produced by these uncorrelated light curves to the original stacked CCF. This is similar to the null hypothesis test used by \citet{Malik2023} to check that the lag recovery is not simply a product of the interaction of the window function with underlying red-noise correlation in the photometric light curves, considering the relatively low sampling density and modest signal-to-noise of our light curves.

\section{Results}\label{sec:results}

We present the results of our stacking analysis for each emission line sample in the OzDES RM sample, and a multi-line analysis of the subsamples for which two emission-lines are present. With the basic assumption that the lag of a source is dependent on the intrinsic AGN luminosity alone, by using the stacking method we are assuming the lags of each source within a bin are similar (in the rest-frame) so that we recover a representative average lag for these AGN. Since the binned sources are at different redshifts, when the light curve data is converted to the rest-frame, different variability and reverberation time-scales are probed by each source through their unique rest-frame observational window function. By stacking we can partially circumvent the usual impact of the sparse sampling (particularly the 7-month seasonal gaps) in the light curves of any one source, as we are combining the CCF's for all sources in a bin. We compare the average lag measurements to the sample of existing individual lag measurements on the respective $R-L$ relations for each line. 

\subsection{H$\beta$}

\begin{figure}
    \centering
    \includegraphics[width=\columnwidth]{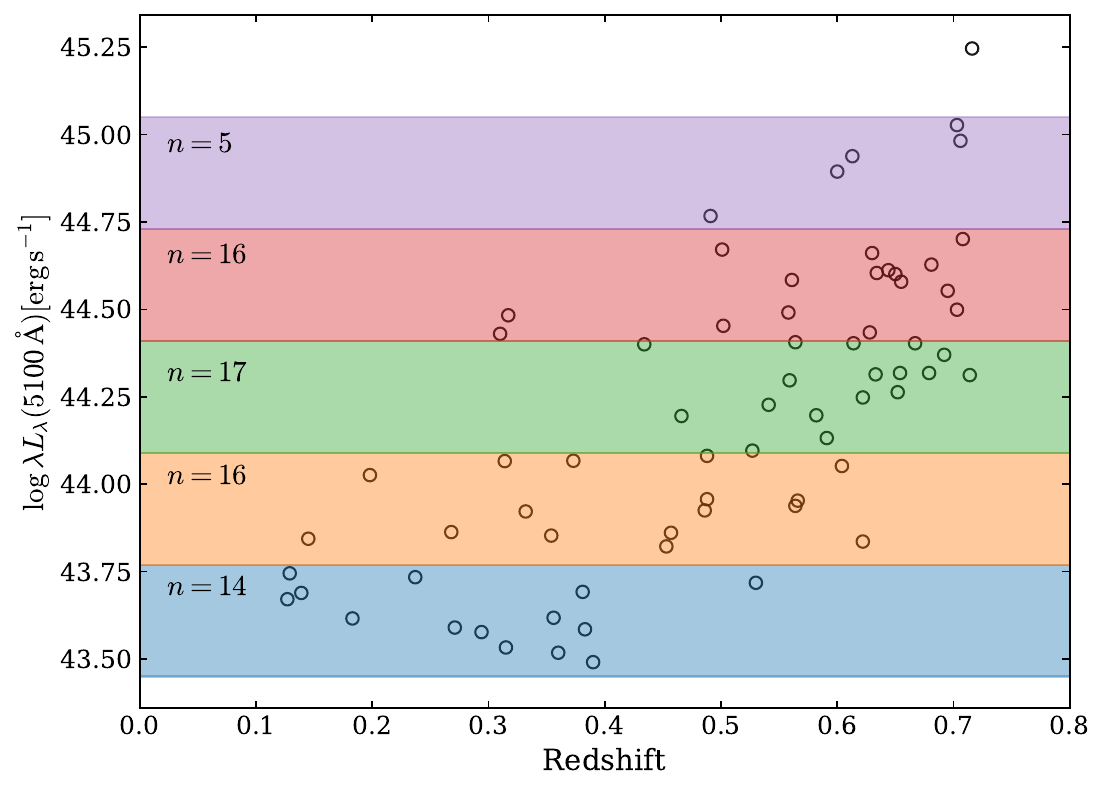}
    \caption{The luminosity bins for the H$\beta$ sample, labelled with the number of binned sources. Within each bin, the standard deviation of the expected lags for the individual sources (measured using the source luminosity and the \citet{Bentz2013} $R-L$ relation for H$\beta$) is $\sim$10\% of the expected mean lag for the binned sample (measured as the mean of the expected lags for the individual sources). }
    \label{fig:Hbeta_bins}
\end{figure}

\begin{figure*}
    \centering
    \includegraphics[width=0.9\textwidth]{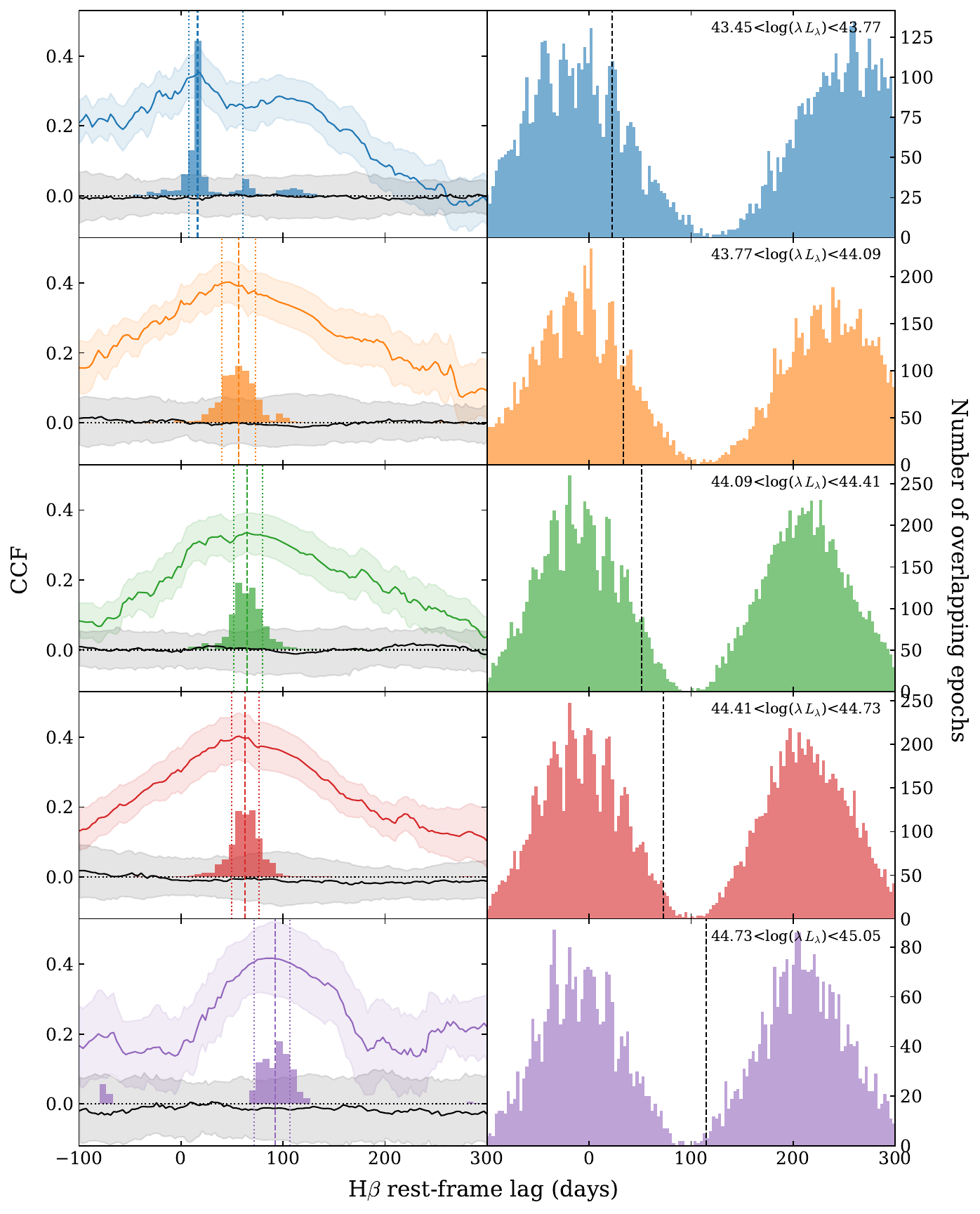}    
    \caption[]{\textit{Left column:} The coloured solid lines are the stacked cross correlation functions (CCF) for each of the five luminosity bins for the H$\beta$ sample. The colours correspond to the respective bins in \autoref{fig:Hbeta_bins}. The 1$\sigma$ scatter of the bootstrapped CCF's is shown by the coloured shaded region. The vertical dashed and dotted lines indicate the recovered average lag and its uncertainty, as measured from the bootstrap distribution (coloured histogram). The black solid line and grey shaded area show the mean and 1$\sigma$ scatter of the CCF's generated using the randomised spectroscopic light curves following the procedure described in \S\ref{sec:method}. \textit{Right column:} The number of overlapping spectroscopic and photometric epochs as a function of time lag, in total for each source in the corresponding bin. The expected mean lag for the bin is indicated by the black dashed line.  } 
    \label{fig:Hbeta_stackedCCF}
\end{figure*}

\begin{figure}
    \centering
    \includegraphics[width=\columnwidth]{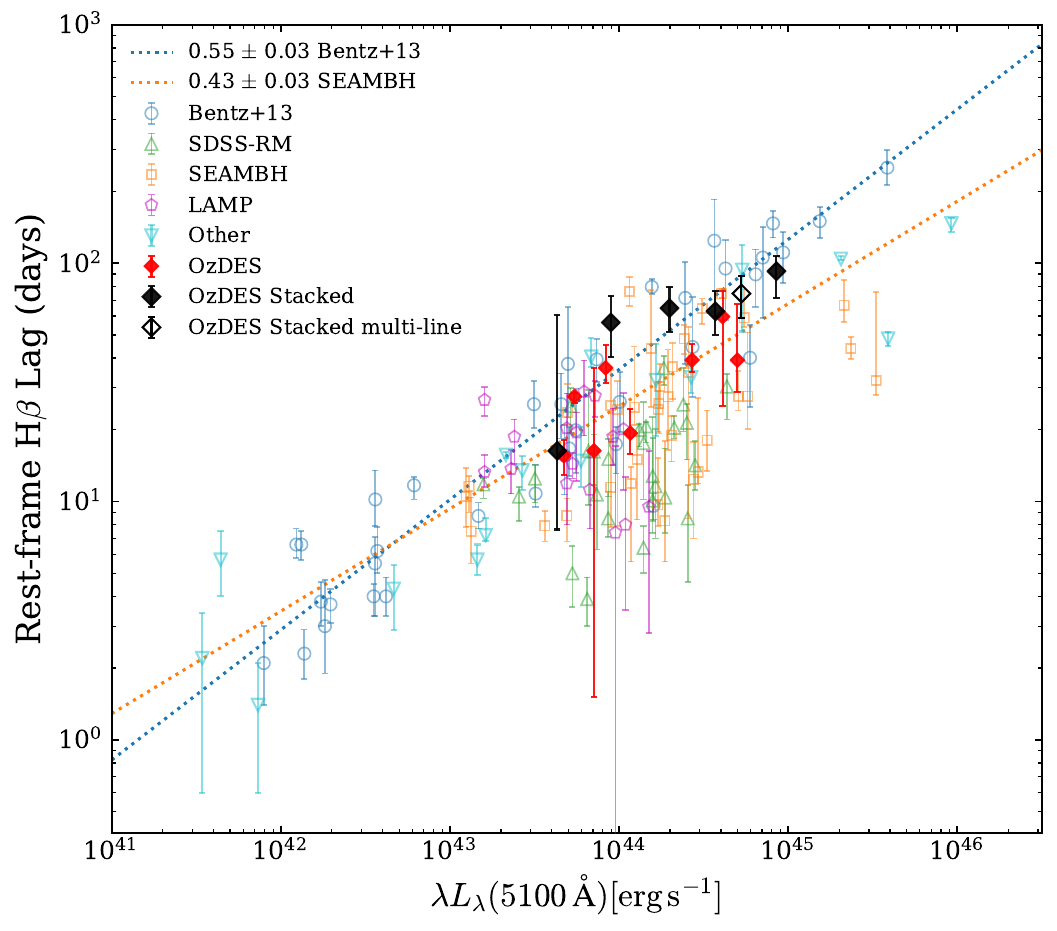} 
    \caption{The Radius-Luminosity relation for H$\beta$ (dotted lines), including the stacked average lag measurements made using the OzDES H$\beta$ sample, and existing individual lag measurements from \citet[][and references therein]{Bentz2013}; SDSS-RM \citep[][quality 4 and 5]{Grier2017}; SEAMBH  \citep{Du2014,Wang2014,Du2015,Du2016,Du2018,Hu2021}; Lick AGN Monitoring Project \citep[LAMP,][]{U2022}; and other measurements from \citet{Bentz2009,Barth2013,Bentz2014,Pei2014,Lu2016,Bentz2016a,Bentz2016b,Fausnaugh2017,Zhang2019,Rakshit2019,Li2021}, of which measurements published before 2019 are compiled by \citet{MartinezAldama2019}. The multi-line average lag measurement is presented in \S\ref{sec:multi}. } 
    \label{fig:Hbeta_RL}
\end{figure}

We divide our sample into five luminosity bins of equal size, as shown in \autoref{fig:Hbeta_bins}. The size of the bins were chosen as to maximise the number of sources in each bin while avoiding introducing a broad underlying distribution in the expected lags. The highest luminosity source was excluded from the analysis for this reason. Although there are a relatively small number of sources stacked in each bin, particularly when compared to stacking done by \citet{Fine2012,Fine2013}, the signal-to-noise of our stacked CCF's are sufficiently high as we have much more light curve data. 

The stacked CCF's for each bin of the H$\beta$ sample are shown in \autoref{fig:Hbeta_stackedCCF}, alongside the total number of overlapping light curve epochs as a function of time-lag for all sources stacked within the bin. As the observation dates of each source in a bin are converted to the rest-frame of the source, and there is a distribution of source redshifts within each bin, the total light curve sampling of the stacked sample begins to in-fill the rest frame observational gap imposed by the observed frame 7-month seasonal gap present in the individual observed light curves.

Comparing the stacked CCF, and its scatter measured from bootstrapping, to the stacked CCF's after light curve randomisation, we see significant correlation signal present. This implies that the signal is not dominated by the correlation of any individual source. In all cases there is one major peak present, however, the lowest luminosity bin (blue) has a flatter CCF. There is limited but non-zero data overlap around the expected mean lags for the two highest luminosity bins, as they coincide with the first seasonal gap in our light curves. However, we recover significant average lags in each of these bins. This demonstrates the ability of stacking to overcome the limitations imposed by sparse sampling, which impede lag recovery for individual sources.

We plot the recovered average lags from each luminosity bin on the $R-L$ relation, as shown in \autoref{fig:Hbeta_RL}. Our stacked measurements are consistent with the \citet{Bentz2013} slope, which agrees with the physically motivated slope of $\sim$0.5. The uncertainty in the average lags are consistent to the uncertainties in the eight lags recovered for individual sources \citep{Malik2023}, and are inconsistent with the distribution of the SDSS-RM measurements \citep{Grier2017,Li2017}, for which shorter lags are recovered. Given the similarities of our programs and sample selection, the reason for this discrepancy is unclear. Although the main difference between the surveys is the baseline and cadence of the data, simulations by \citet{FonsecaAlvarez2020} and \citet{Malik2022} find that this does not bias the lag recovery of SDSS-RM to shorter lags, or OzDES to longer lags. For our composite lags we bin across our entire H$\beta$ sample, and do not reject any sources based on light curve signal-to-noise, or any other criteria.

As a test, we repeated the stacking after excluding all the objects with individual recovered lags \citep{Malik2023}. We show the results of this test in \S\ref{sec:appendixB}. Although the average lag uncertainties increase after the exclusion, the lags remain in close agreement with those from the original analysis. 

\subsection{Mg\,\textsc{ii}}

\begin{figure}
    \centering
    \includegraphics[width=\columnwidth]{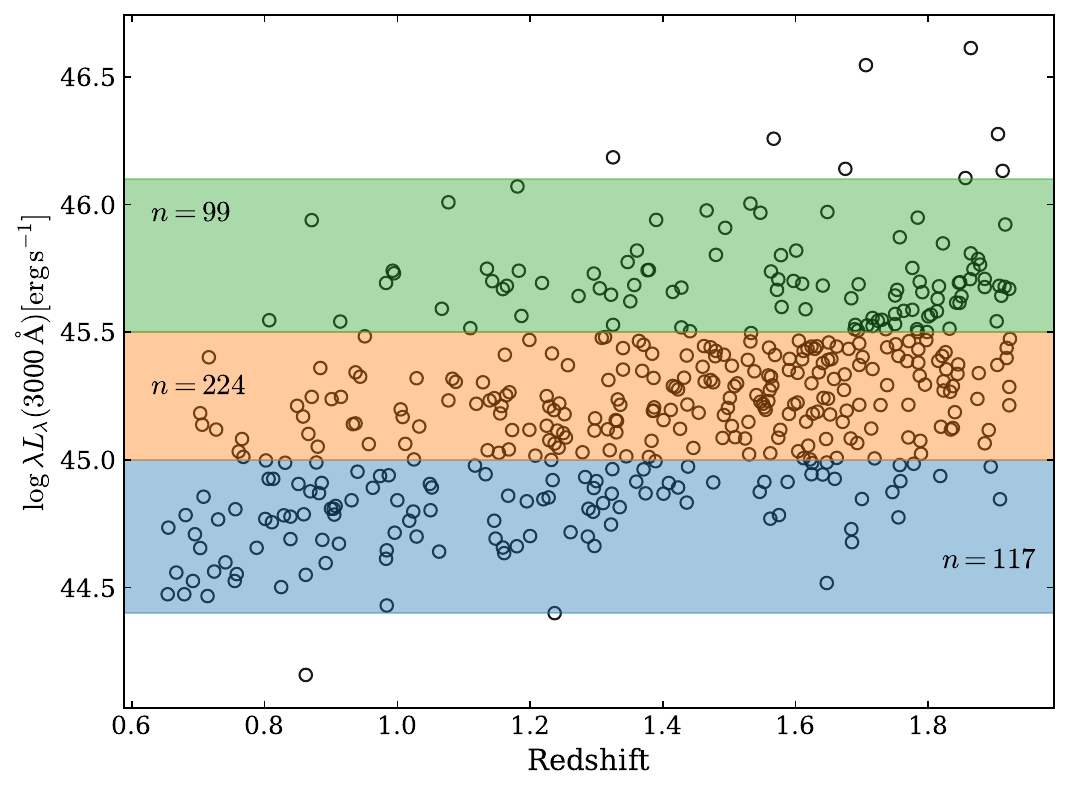}
    \caption{The luminosity bins for the Mg\,\textsc{ii} sample, labelled with the number of binned sources. Within each bin, the standard deviation of the expected lags for the individual sources (measured using the source luminosity and the \citet{Trakhtenbrot2012} $R-L$ relation for Mg\,\textsc{ii}) is $\sim$20\% of the expected mean lag for the binned sample.}
    \label{fig:MgII_bins}
\end{figure}

\begin{figure*}
    \centering
    \includegraphics[width=0.85\textwidth]{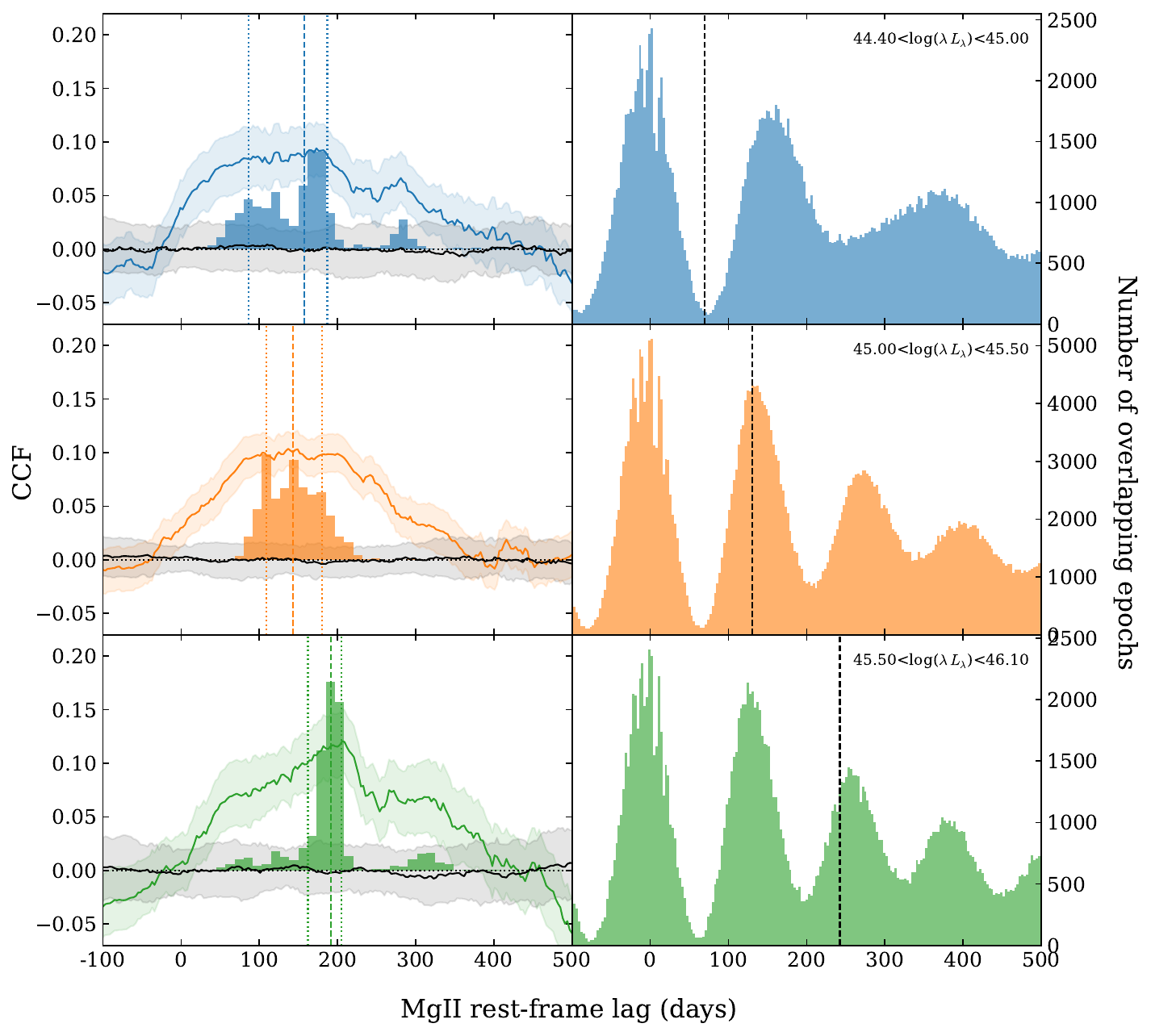} 
    \caption[]{\textit{Left column:} The coloured solid lines are the stacked cross correlation functions (CCF) for each of the three luminosity bins for the Mg\,\textsc{ii} sample. The colours correspond to the respective bins in \autoref{fig:MgII_bins}. The 1$\sigma$ scatter of the bootstrapped CCF's is shown by the coloured shaded region. The vertical dashed and dotted lines indicate the recovered average lag and its uncertainty, as measured from the bootstrap distribution (coloured histogram). The black solid line and grey shaded area show the mean and 1$\sigma$ scatter of the CCF's generated using the randomised spectroscopic light curves following the procedure described in the text. \textit{Right column:} The number of overlapping spectroscopic and photometric epochs as a function of time lag, in total for each source in the corresponding bin. The expected mean lag for the bin is indicated by the black dashed line.} 
    \label{fig:MgII_stackedCCF}
\end{figure*}

\begin{figure}
    \centering
    \includegraphics[width=\columnwidth]{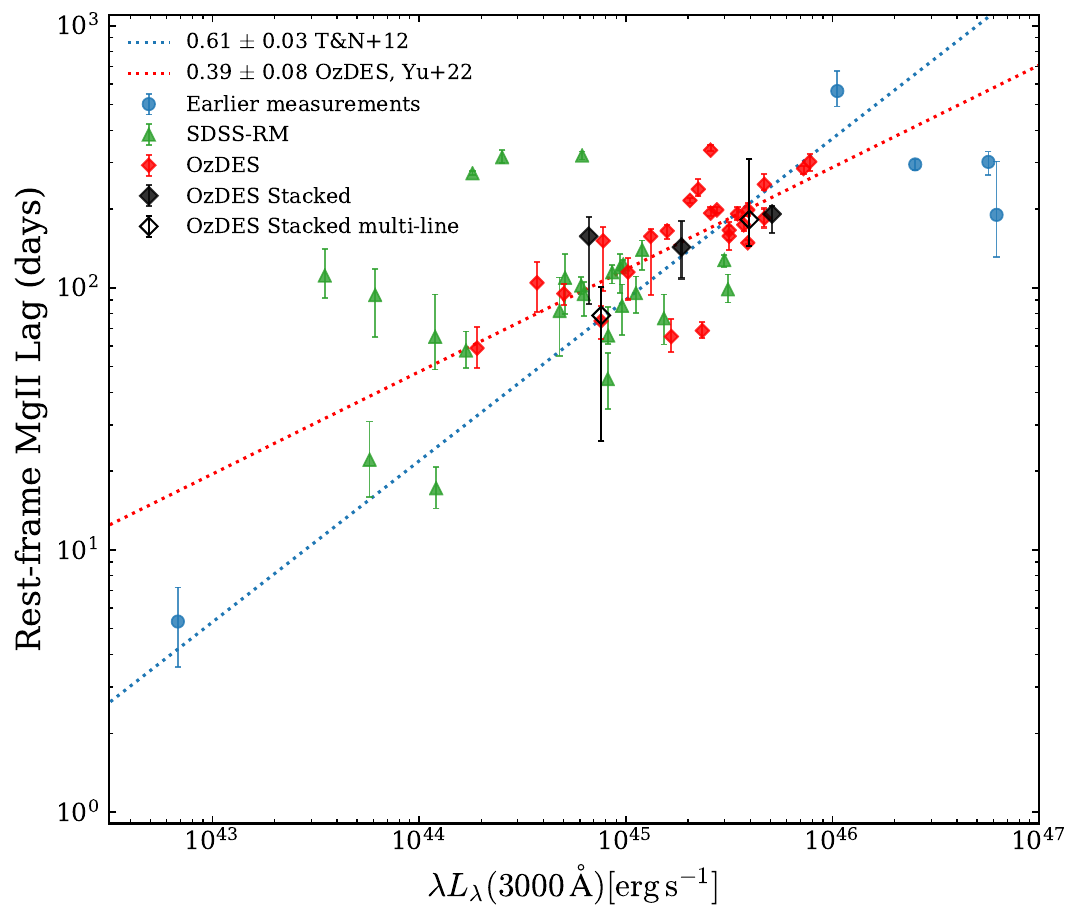}
    \caption{Radius-Luminosity relation for Mg\,\textsc{ii}, with existing individual lag measurements from \citet{Metzroth2006,Lira2018,Czerny2019,Zajacek2020,Zajacek2021} and SDSS-RM \citep[][gold sample]{Homayouni2020}, along with the OzDES individual measurements \citep{Yu2023}, and our stacked average lag measurements. The multi-line average lag measurements are presented in \S\ref{sec:multi}.}
    \label{fig:MgII_RL}
\end{figure}

The three luminosity bins used for the Mg\,\textsc{ii} sample are shown in \autoref{fig:MgII_bins}. The lowest and highest luminosity bins are slightly wider to include sources close to the edge of the bin. Ten sources were excluded from the analysis to optimise the bin densities and reduced smoothing of the stacked CCF's. Larger luminosity bins were required to achieve adequate signal-to-noise in the stacked CCF's for this sample.

We present the stacked CCF's in \autoref{fig:MgII_stackedCCF}. The strength of the correlation signals in each bin are lower than for the H$\beta$ sample, however, comparing to the CCF's produced using randomised light curves we can see there is significant signal present. The signal-to-noise of the Mg\,\textsc{ii} light-curve input data is lower than for the H$\beta$ sample, as the Mg\,\textsc{ii} sample is fainter and requires subtraction of Fe\,\textsc{ii} from the emission-line \citep{Yu2021,Yu2023}. However, we have 450 AGN in the Mg\,\textsc{ii} sample, and are therefore stacking many more sources in each bin. Since sample size is not the limiting factor in this case, the lower signal-to-noise of the Mg\,\textsc{ii} line light curves must be producing weaker stacked cross correlation signals. In the lowest luminosity bin, the signal is flat and no clear peak is present. For the other two bins a dominant peak is present, and the bootstrap distributions are adequately constrained to recover average lags. 

We plot the recovered average lags from each luminosity bin on the Mg\,\textsc{ii} $R-L$ relation in \autoref{fig:MgII_RL}, along with the 25 individual measurements made with this sample by \citet{Yu2023}. The average lags are in agreement with the recent individual measurements from \citet{Homayouni2020} and \citet{Yu2023}. There is no clear progression to longer lags with higher luminosities given that the average lags for the lowest and intermediate luminosity bins are not particularly well constrained. As shown in \autoref{fig:MgII_stackedCCF}, there is limited data coverage over the expected mean lags for the  lowest luminosity bin (blue), which coincides with the first seasonal gap in our light curves. The time-dilation distribution over the binned sample is not sufficient to `fill in' the short timescales, but it is able to sufficiently bridge the second seasonal gap. This could explain why the uncertainty on the recovered average lag for the lowest luminosity bin is larger than the uncertainties for the intermediate and highest luminosity bins. The average lags for the three luminosity bins are formally consistent with both the steeper $R-L$ relation measured by \citet{Trakhtenbrot2012}, and the shallower relation recently constrained by \citet{Yu2023}, however, the better constrained mean lag for the highest luminosity bin is only consistent with the shallower relation.

\subsection{C\,\textsc{iv}}

\begin{figure}
    \centering
    \includegraphics[width=\columnwidth]{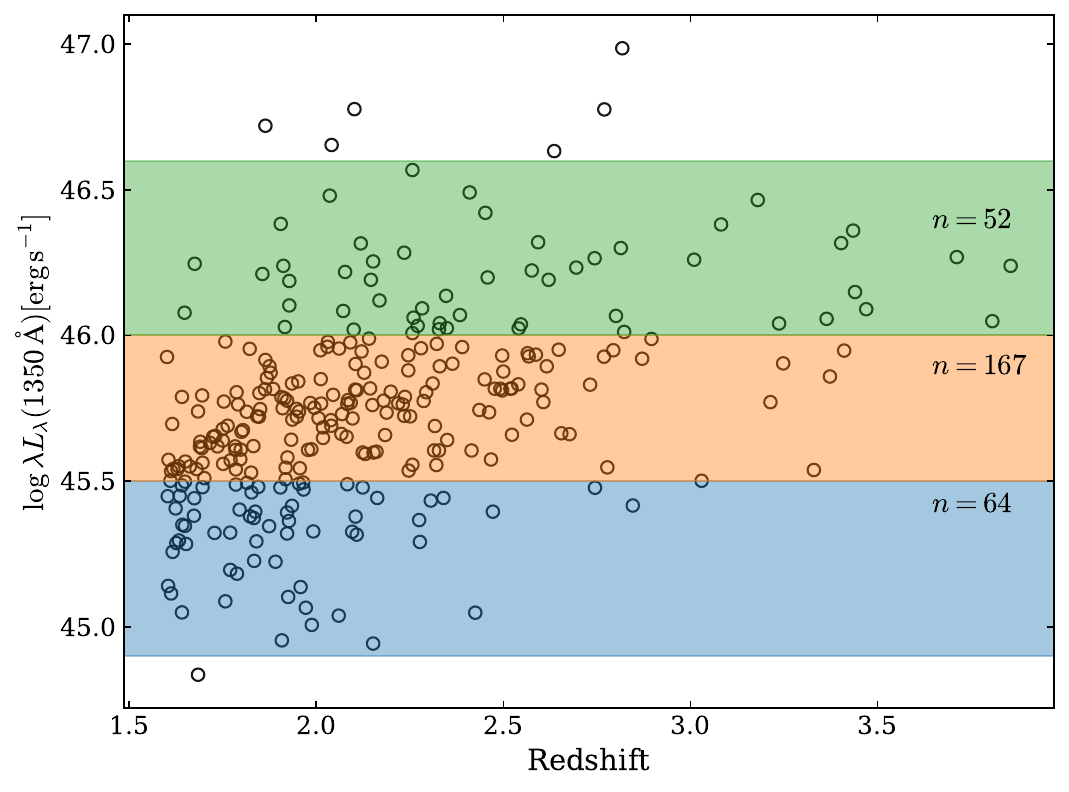}
    \caption{The luminosity bins for the C\,\textsc{iv} sample, labelled with the number of binned sources. Within each bin, the standard deviation of the expected lags for the individual sources (measured using the source luminosity and the \citet{Hoormann2019} $R-L$ relation for C\,\textsc{iv}) is $\sim$15\% of the expected mean lag for the binned sample.}
    \label{fig:CIV_bins}
\end{figure}

\begin{figure*}
    \centering
    \includegraphics[width=0.85\textwidth]{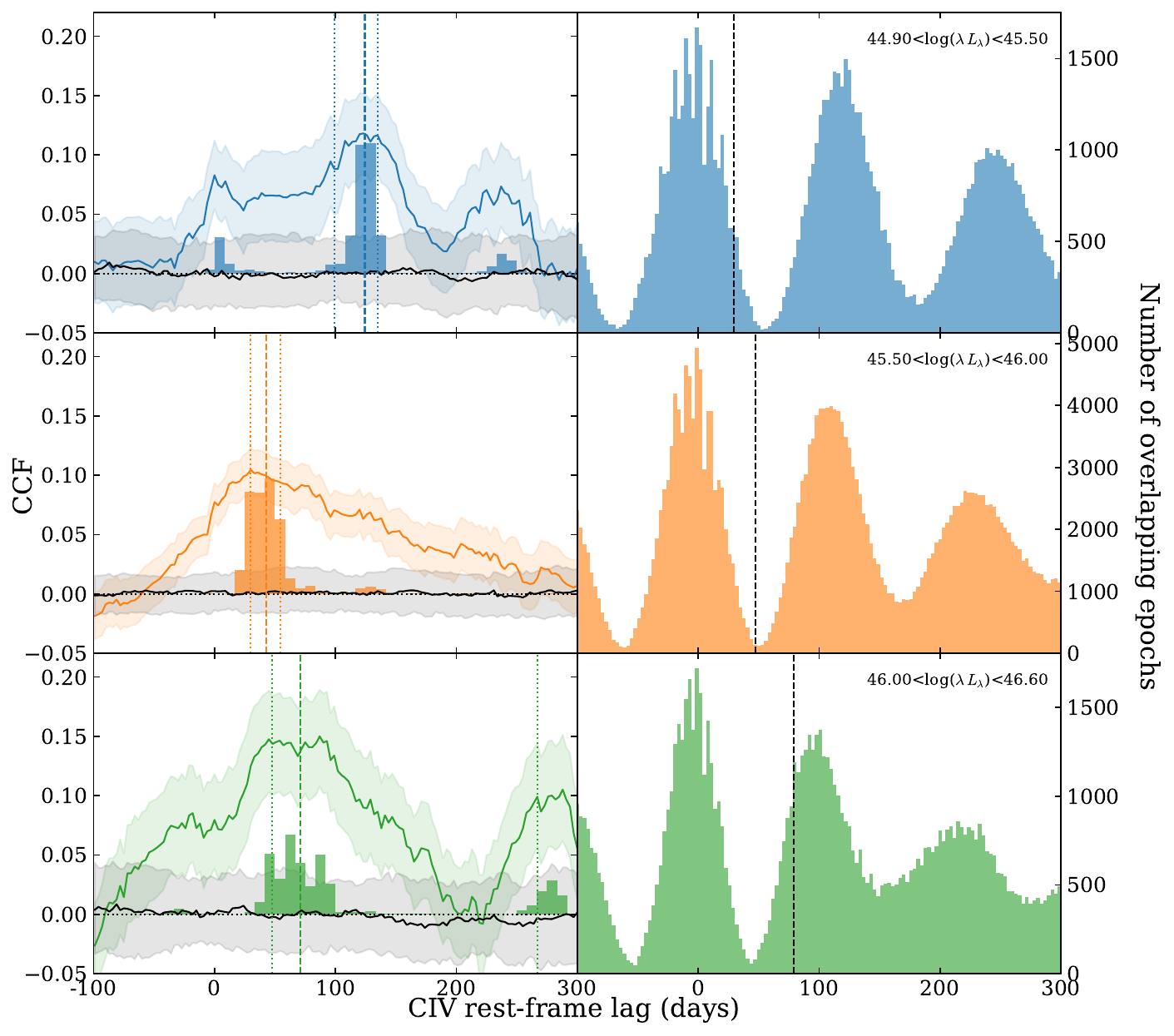} 
    \caption[]{\textit{Left column:} The coloured solid lines are the stacked cross correlation functions (CCF) for each of the three luminosity bins for the C\,\textsc{iv} sample. The colours correspond to the respective bins in \autoref{fig:CIV_bins}. The 1$\sigma$ scatter of the bootstrapped CCF's is shown by the coloured shaded region. The vertical dashed and dotted lines indicate the recovered average lag and its uncertainty, as measured from the bootstrap distribution (coloured histogram). The black solid line and grey shaded area show the mean and 1$\sigma$ scatter of the CCF's generated using the randomised spectroscopic light curves following the procedure described in the text. \textit{Right column:} The number of overlapping spectroscopic and photometric epochs as a function of time lag, in total for each source in the corresponding bin. The expected mean lag for the bin is indicated by the black dashed line.} 
    \label{fig:CIV_stackedCCF}
\end{figure*}

\begin{figure}
    \centering
    \includegraphics[width=\columnwidth]{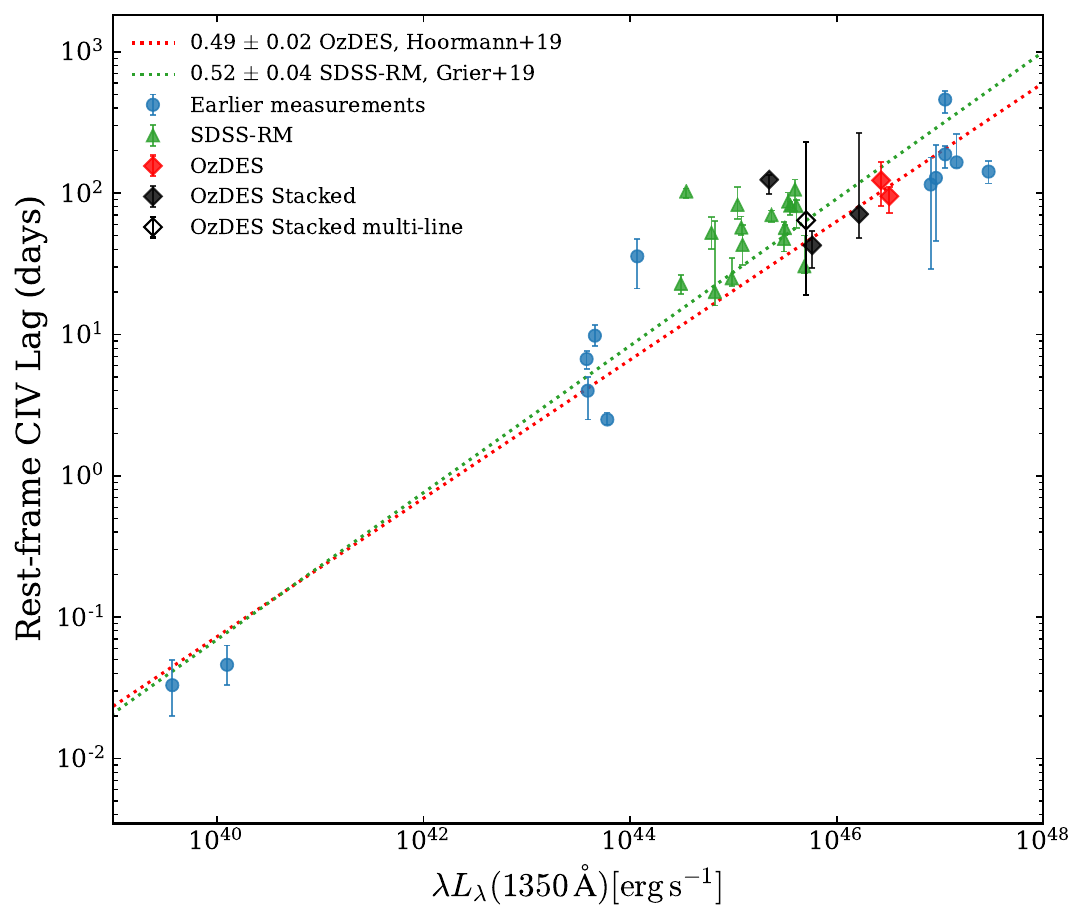}
    \caption{Radius-Luminosity relation for C\,\textsc{iv}, with our stacked average lags and existing individual lag measurements from \citet[][and references therein]{Peterson2005}; \citet{Kaspi2007,Trevese2014,Lira2018}, and SDSS-RM \citep[][gold sample]{Grier2019,Shen2019}, \citet{Kaspi2021}, along with the OzDES individual measurements \citep{Hoormann2019}. The multi-line average lag measurement is presented in \S\ref{sec:multi}. }
    \label{fig:CIV_RL}
\end{figure}

As for the previous samples, we present the luminosity bins for the C\,\textsc{iv} sample in \autoref{fig:CIV_bins}, and the stacked CCF results in \autoref{fig:CIV_stackedCCF}. As required for Mg\,\textsc{ii}, larger luminosity bins were necessary to achieve sufficient signal-to-noise after stacking, and nine sources were excluded from the analysis to avoid overly wide bins in luminosity. The OzDES C\,\textsc{iv} sample has lower signal-to-noise than the H$\beta$ and Mg\,\textsc{ii} samples, as the AGN are faint. The bootstrap distributions are not well constrained for the lowest or highest luminosity bins. However, \citet{Li2017} found that the mean recovered lag remains stable when the light curve signal-to-noise is degraded, although the uncertainty increases proportionally with the decline. Therefore we continue to recover the average lags and compare with individual source measurements. 

We plot the stacked average lags from the C\,\textsc{iv} sample alongside existing measurements from the literature in \autoref{fig:CIV_RL}. The average lags for the intermediate and highest luminosity bins are in agreement with the $R-L$ relations constrained by \citet{Hoormann2019} and \citet{Grier2019}, although the uncertainty on the average lag for the highest luminosity bin is large. As discussed for the Mg\,\textsc{ii} sample, the time dilation over the binned samples does not sufficiently bridge the first of the 7-month seasonal gaps in our light curves. With the shorter expected lags for C\,\textsc{iv}, the average lags for the lowest and intermediate luminosity bins coincides with this gap. In addition to the lower signal-to-noise of the light curves, this may be contributing to the poorer quality of the stacked CCF's. It is unclear why the average lag recovered for the lowest luminosity bin is much longer than expected.

\subsection{Multi-line measurements}\label{sec:multi}

Both the H$\beta$ and Mg\,\textsc{ii} lines are visible for 13 AGN, and Mg\,\textsc{ii} and C\,\textsc{iv} for 106 AGN (see \autoref{fig:ozdes_AGN}). We attempt to recover average lags independently with each line, in order to compare the lag ratios to investigate the ionisation stratification of the BLR.  

\begin{figure}
    \centering
    \includegraphics[width=\columnwidth]{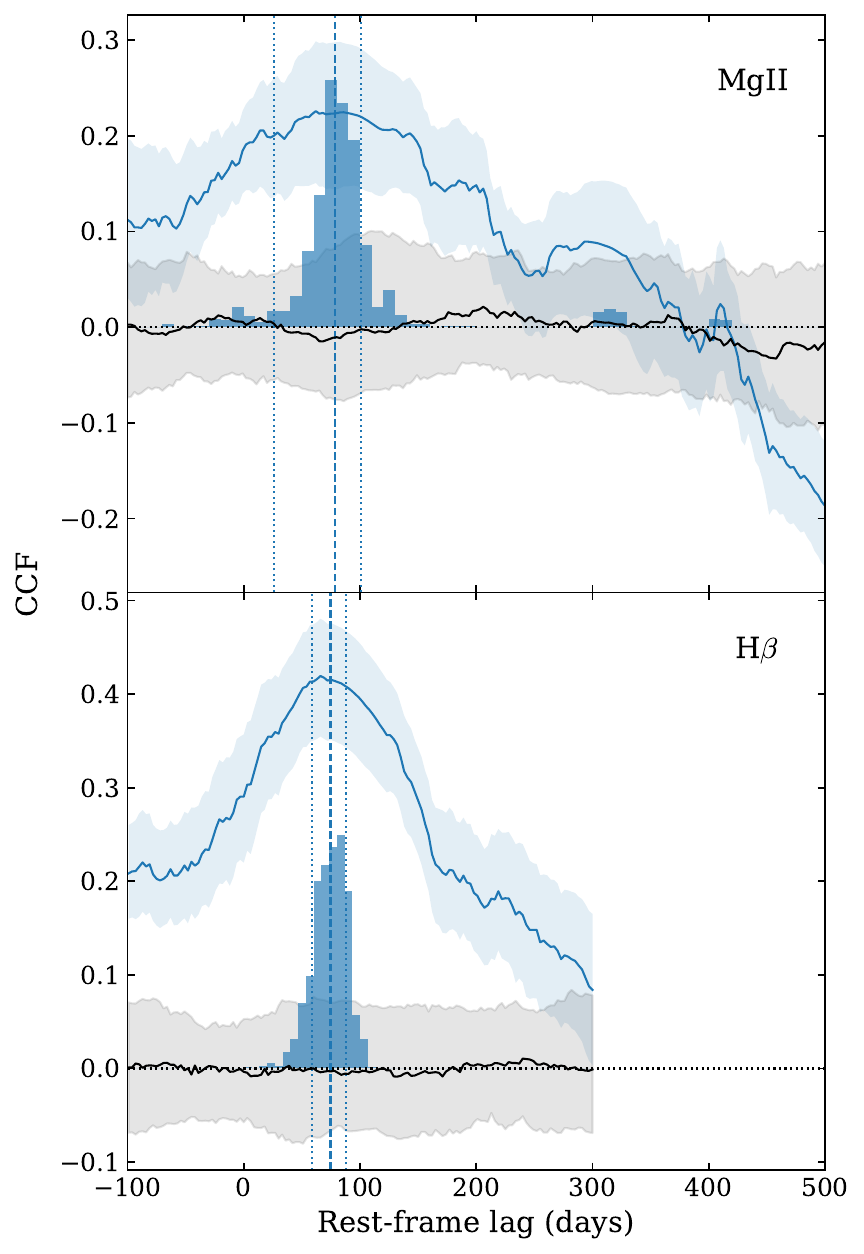}
    \caption{The stacked cross correlation functions (CCF) for the multi-line sample with both H$\beta$ and Mg\,\textsc{ii}, comprising 13 AGN. }
    \label{fig:H_M_CCF}
\end{figure}

We repeated the stacking procedure for the sample of 13 AGN with the H$\beta$ and Mg\,\textsc{ii} light curves. As there are few sources we do not bin them by luminosity. We present the stacked CCF's for each line in \autoref{fig:H_M_CCF}. We measure an average lag of 75$^{+14}_{-16}$ days for the H$\beta$ sample, and an average lag of 79$^{+22}_{-53}$ days for the Mg\,\textsc{ii} sample. We formally recover a Mg\,\textsc{ii} to H$\beta$ lag ratio of 1.05$\pm0.54$, however the inherent uncertainty is significant due to the large uncertainty of the Mg\,\textsc{ii} average lag. This result is broadly consistent with the expectation that these two BLRs are approximately cospatial. This ratio is consistent with previous multi-line measurements made by \citet{Homayouni2020}, who found that Mg\,\textsc{ii} is emitted from a similar or slightly larger region than H$\beta$ in several individual sources, as well as \citet{Clavel1991} and \citet{Czerny2019}. 

\begin{figure}
    \centering
    \includegraphics[width=\columnwidth]{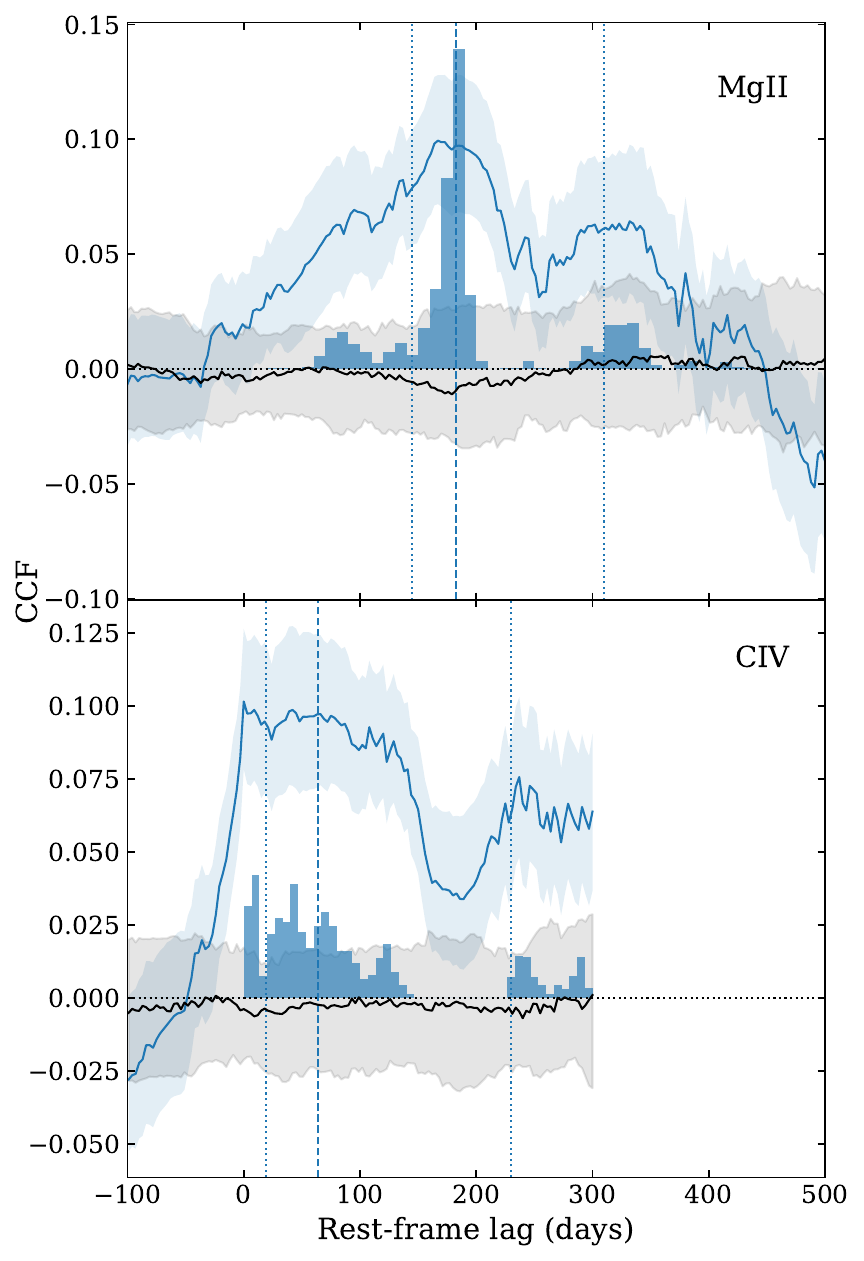}
    \caption{The stacked cross correlation functions (CCF) for the multi-line sample with both Mg\,\textsc{ii} and C\,\textsc{iv}, comprising 106 AGN. }
    \label{fig:M_C_CCF}
\end{figure}

We repeated this for the sample of 106 AGN with Mg\,\textsc{ii} and C\,\textsc{iv}, and present the stacked CCF's in \autoref{fig:M_C_CCF}. The signal-to-noise was insufficient to divide the sample into two luminosity bins. We measure an average Mg\,\textsc{ii} lag of 182$^{+128}_{-37}$ days, and an average C\,\textsc{iv} lag of 64$^{+165}_{-45}$ days. We formally recover a Mg\,\textsc{ii} to C\,\textsc{iv} lag ratio of 2.84$\pm4.84$, however this is poorly (if at all) constrained given that the average lags are not well constrained (particularly for the CIV sample). The average lags are broadly consistent with the BLR stratification model, and the multi-line comparison made for a single source by \citet{Homayouni2020}. 

We include the average lags recovered from the multi-line samples on the respective $R-L$ plots presented in \autoref{fig:Hbeta_RL}, \autoref{fig:MgII_RL} and \autoref{fig:CIV_RL}. We also provide all average lags recovered in this work from each multi-line and emission line sample in \autoref{tab:stacked_lags}.

\renewcommand{\arraystretch}{1.3}
\begin{table}
    \centering
\caption{The average lags for each luminosity bin or multi-line sample, for each emission-line sample from the OzDES RM Program. Luminosities are given in erg s$^{-1}$, and average lags are in the rest frame.}
    \begin{tabular}{cc}
    \hline
    \multicolumn{2}{c}{H$\beta$}                                                           \\ \hline
    \multicolumn{1}{c}{Binned sample} & \multicolumn{1}{c}{Average lag (days)} \\ \hline
    43.45$<$log($\lambda L_\lambda)<$43.77   &   $16^{+44}_{-9}$ \\
    43.77$<$log($\lambda L_\lambda)<$44.09   &   $56^{+16}_{-16}$ \\
    44.09$<$log($\lambda L_\lambda)<$44.41   &   $65^{+15}_{-13}$ \\
    44.41$<$log($\lambda L_\lambda)<$44.73   &   $63^{+13}_{-13}$ \\
    44.73$<$log($\lambda L_\lambda)<$45.05   &   $93^{+14}_{-21}$ \\
    multi-line H$\beta$ + Mg\,\textsc{ii} &   $75^{+14}_{-16}$ \\ \hline
    \multicolumn{2}{c}{Mg\,\textsc{ii}}                               \\ \hline
    Binned sample                    & Average lag (days)                     \\ \hline
    44.40$<$log($\lambda L_\lambda)<$45.00   &   $157^{+30}_{-71}$ \\
    45.00$<$log($\lambda L_\lambda)<$45.50   &   $143^{+38}_{-34}$ \\
    45.50$<$log($\lambda L_\lambda)<$46.10   &   $191^{+14}_{-29}$ \\
    multi-line Mg\,\textsc{ii} + H$\beta$  &   $79^{+22}_{-53}$ \\
    multi-line Mg\,\textsc{ii} + C\,\textsc{iv}   &   $182^{+128}_{-37}$ \\ \hline
    \multicolumn{2}{c}{C\,\textsc{iv}}                                \\ \hline
    Binned sample                    & Average lag (days)                     \\ \hline
    44.90$<$log($\lambda L_\lambda)<$45.50   &   $124^{+11}_{-25}$ \\
    45.50$<$log($\lambda L_\lambda)<$46.00   &   $43^{+12}_{-13}$ \\
    46.00$<$log($\lambda L_\lambda)<$46.60   &   $71^{+196}_{-23}$ \\
    multi-line C\,\textsc{iv} + Mg\,\textsc{ii}  &   $64^{+165}_{-45}$          
    \end{tabular}
	\label{tab:stacked_lags}
\end{table}

\section{Summary}\label{sec:summary}

We use the stacking technique developed by \citet{Fine2012,Fine2013} to measure average lags in luminosity bins for the H$\beta$, Mg\,\textsc{ii} and C\,\textsc{iv} samples from the OzDES Reverberation Mapping Program. By utilising the bulk of our sample to recover composite lags, we avoid the potential selection biases in the lag recovery for individual sources. We successfully recover significant cross-correlation signals for each emission-line sample:
\begin{itemize}
    \item The average lags from each luminosity bin of the H$\beta$ sample are consistent with the $R-L$ relation constrained by \citet{Bentz2013}, and the size of the uncertainties on the average lags are on par with that of individual measurements, despite the relatively small number of sources stacked in each bin. This provides confidence in the individual measurements, and demonstrates the potential for stacked RM analyses to improve upon constraints which have thus far been made with individual source measurements alone.
    \item For Mg\,\textsc{ii} and C\,\textsc{iv}, the stacked cross-correlations are weaker, but still present above the correlation signal generated using randomised light curves. Further data or larger samples are required to recover significant average lags for these samples across each luminosity bin. 
    \item From our multi-line analysis, we measure a Mg\,\textsc{ii} to H$\beta$ lag ratio that is consistent with earlier findings that the size of each of these line-emitting regions is similar. Our average lags for the Mg\,\textsc{ii} and C\,\textsc{iv} multi-line sample are not well-constrained due to the limited sample size, however, the lag ratio we recover is largely consistent with the BLR ionisation stratification model. 
\end{itemize}

Stacking can be applied beyond RM-specific surveys. It can be done with just a few spectroscopic epochs, using large AGN samples, provided the continuum behaviour is well sampled. With LSST forthcoming, high cadence photometry of the deep-drilling fields will yield quality continuum light curves for tens of thousands of AGN. The SDSS-V Black Hole Mapper (BHM) will be spectroscopically following these fields. In addition to their dedicated RM survey of $\sim1,000$ AGN, SDSS-V BHM will monitor 25,000 AGN over multiple epochs, which will be combined with earlier SDSS spectra \citep{Kollmeier2017}. The Time Domain Extragalactic Survey (TiDES) will also be following up these fields to conduct an RM survey of $\sim700$ AGN up to $z\sim2.5$ \citep{Swann2019}. Stacking analyses of these future, large samples has promise to significantly extend the reverberation mapping results from these projects. Although the improved signal-to-noise and survey sampling of these future programs are expected to yield an increased number of higher quality individual lag measurements than the first generation of multi-object RM surveys, these datasets can benefit from the ability of stacking to alleviate the impact of the unavoidable seasonal gaps on lag recovery \citep{Malik2022}. Stacking also presents an opportunity to combine all large time-domain datasets to recover average lags that are potentially more robust than individual lags, which remain challenging to recover reliably, particularly at high redshift.

\section*{Acknowledgements}

We thank the anonymous referee for their comments that improved the paper. UM and AP are supported by the Australian Government Research Training Program (RTP) Scholarship. PM and ZY are supported in part by the United States National Science Foundation under Grant No. 161553 to PM. PM also acknowledges support from the United States Department of Energy, Office of High Energy Physics under Award Number DE-SC-0011726. TMD is supported by an Australian Research Council Laureate Fellowship (project number FL180100168). 

We acknowledge parts of this research were carried out on the traditional lands of the Ngunnawal and Ngambri peoples. This work makes use of data acquired at the Anglo-Australian Telescope, under program A/2013B/012. We acknowledge the Gamilaraay people as the traditional owners of the land on which the AAT stands. We pay our respects to their elders past and present.

This analysis used \texttt{NumPy} \citep{harris2020array}, \texttt{Astropy} \citep{astropy:2013, astropy:2018}, and \texttt{SciPy} \citep{2020SciPy-NMeth}. Plots were made using \texttt{Matplotlib} \citep{Hunter:2007}. This work has made use of the SAO/NASA Astrophysics Data System Bibliographic Services.

This paper has gone through internal review by the DES collaboration. Funding for the DES Projects has been provided by the U.S. Department of Energy, the U.S. National Science Foundation, the Ministry of Science and Education of Spain, 
the Science and Technology Facilities Council of the United Kingdom, the Higher Education Funding Council for England, the National Center for Supercomputing 
Applications at the University of Illinois at Urbana-Champaign, the Kavli Institute of Cosmological Physics at the University of Chicago, 
the Center for Cosmology and Astro-Particle Physics at the Ohio State University,
the Mitchell Institute for Fundamental Physics and Astronomy at Texas A\&M University, Financiadora de Estudos e Projetos, 
Funda{\c c}{\~a}o Carlos Chagas Filho de Amparo {\`a} Pesquisa do Estado do Rio de Janeiro, Conselho Nacional de Desenvolvimento Cient{\'i}fico e Tecnol{\'o}gico and 
the Minist{\'e}rio da Ci{\^e}ncia, Tecnologia e Inova{\c c}{\~a}o, the Deutsche Forschungsgemeinschaft and the Collaborating Institutions in the Dark Energy Survey. 

The Collaborating Institutions are Argonne National Laboratory, the University of California at Santa Cruz, the University of Cambridge, Centro de Investigaciones Energ{\'e}ticas, 
Medioambientales y Tecnol{\'o}gicas-Madrid, the University of Chicago, University College London, the DES-Brazil Consortium, the University of Edinburgh, 
the Eidgen{\"o}ssische Technische Hochschule (ETH) Z{\"u}rich, 
Fermi National Accelerator Laboratory, the University of Illinois at Urbana-Champaign, the Institut de Ci{\`e}ncies de l'Espai (IEEC/CSIC), 
the Institut de F{\'i}sica d'Altes Energies, Lawrence Berkeley National Laboratory, the Ludwig-Maximilians Universit{\"a}t M{\"u}nchen and the associated Excellence Cluster Universe, 
the University of Michigan, NSF's NOIRLab, the University of Nottingham, The Ohio State University, the University of Pennsylvania, the University of Portsmouth, 
SLAC National Accelerator Laboratory, Stanford University, the University of Sussex, Texas A\&M University, and the OzDES Membership Consortium.

Based in part on observations at Cerro Tololo Inter-American Observatory at NSF's NOIRLab (NOIRLab Prop. ID 2012B-0001; PI: J. Frieman), which is managed by the Association of Universities for Research in Astronomy (AURA) under a cooperative agreement with the National Science Foundation.

The DES data management system is supported by the National Science Foundation under Grant Numbers AST-1138766 and AST-1536171.
The DES participants from Spanish institutions are partially supported by MICINN under grants ESP2017-89838, PGC2018-094773, PGC2018-102021, SEV-2016-0588, SEV-2016-0597, and MDM-2015-0509, some of which include ERDF funds from the European Union. IFAE is partially funded by the CERCA program of the Generalitat de Catalunya.
Research leading to these results has received funding from the European Research
Council under the European Union's Seventh Framework Program (FP7/2007-2013) including ERC grant agreements 240672, 291329, and 306478.
We  acknowledge support from the Brazilian Instituto Nacional de Ci\^encia
e Tecnologia (INCT) do e-Universo (CNPq grant 465376/2014-2).

This manuscript has been authored by Fermi Research Alliance, LLC under Contract No. DE-AC02-07CH11359 with the U.S. Department of Energy, Office of Science, Office of High Energy Physics.

\section*{Data Availability}
The underlying DES and OzDES data are available in \citet{Abbott2021} and \citet{Lidman2020}. The final light curve data for the full OzDES RM data set will be made available in a future OzDES paper.
 




\bibliographystyle{mnras}
\bibliography{refs} 



\appendix

\section{Source properties}\label{sec:appendixA}

The source properties are given for the H$\beta$ sample of 69 AGN in \autoref{tab:AGNpropH}, the Mg\,\textsc{ii} sample of 450 AGN in \autoref{tab:AGNpropM}, and the C\,\textsc{iv} sample of 290 AGN in \autoref{tab:AGNpropC}. 

\renewcommand{\arraystretch}{1.0}
\begin{table}
\caption{Properties for our OzDES H$\beta$ stacking sample. Columns left to right: DES name (J2000), redshift, $r$-band apparent AB magnitude, monochromatic luminosity at 5100\AA. The superscript $a$ flags sources which also have Mg\,\textsc{ii} data. }
\begin{tabular}{llll}
DES ID & $z$ & $m_r$ & $\log({\lambda}L_{5100})$ (erg\,s$^{-1}$)  \\ \hline
DES J002802.42-424913.52 & 0.127 & 17.88 & 43.67 \\
DES J033633.16-284027.18 & 0.129 & 17.73 & 43.74 \\
DES J022024.92-061731.51 & 0.139 & 18.04 & 43.69 \\
DES J003515.60-433357.64 & 0.145 & 17.75 & 43.84 \\
DES J003245.59-421441.67 & 0.183 & 18.85 & 43.62 \\
DES J025007.02+002525.49 & 0.198 & 18.02 & 44.03 \\
DES J024347.34-005354.84 & 0.237 & 19.19 & 43.73 \\
DES J024340.97-002601.16 & 0.268 & 19.16 & 43.86 \\
DES J025021.72-005413.02 & 0.271 & 19.87 & 43.59 \\
DES J025211.61-003629.35 & 0.294 & 20.10 & 43.58 \\
DES J034028.46-292902.41 & 0.310 & 18.09 & 44.43 \\
DES J022249.67-051453.01 & 0.314 & 19.03 & 44.07 \\
DES J025240.59-002117.04 & 0.315 & 20.37 & 43.53 \\
DES J022851.50-051223.00 & 0.317 & 18.01 & 44.48 \\
DES J003954.13-440509.97 & 0.332 & 19.53 & 43.92 \\
DES J022330.16-054758.06 & 0.354 & 19.87 & 43.85 \\
DES J024325.53-000412.67 & 0.356 & 20.47 & 43.62 \\
DES J024902.03-004322.43 & 0.360 & 20.75 & 43.52 \\
DES J024320.36-001825.41 & 0.373 & 19.47 & 44.07 \\
DES J024519.82-010245.63 & 0.381 & 20.46 & 43.69 \\
DES J024225.86-004142.50 & 0.383 & 20.74 & 43.59 \\
DES J003622.82-424759.11 & 0.390 & 21.02 & 43.49 \\
DES J004009.06-431255.29 & 0.434 & 19.04 & 44.40 \\
DES J003834.47-433807.12 & 0.453 & 20.61 & 43.82 \\
DES J025042.90-004138.74 & 0.457 & 20.54 & 43.86 \\
DES J022258.90-045852.28 & 0.466 & 19.77 & 44.20 \\
DES J024712.89-011106.21 & 0.486 & 20.56 & 43.92 \\
DES J024211.93-010959.69 & 0.488 & 20.18 & 44.08 \\
DES J024538.29-004705.32 & 0.488 & 20.49 & 43.96 \\
DES J003017.47-422446.39 & 0.491 & 18.48 & 44.77 \\
DES J022133.82-054842.69 & 0.501 & 18.77 & 44.67 \\
DES J024643.04-013149.55 & 0.502 & 19.32 & 44.45 \\
DES J033002.93-273248.31 & 0.527 & 20.33 & 44.10 \\
DES J003552.21-423352.14 & 0.530 & 21.29 & 43.72 \\
DES J022019.61-060729.77 & 0.541 & 20.07 & 44.23 \\
DES J021910.57-055114.76 & 0.558 & 19.49 & 44.49 \\
DES J025119.79-004831.62 & 0.559 & 19.98 & 44.30 \\
DES J033758.00-294618.46 & 0.561 & 19.27 & 44.58 \\
DES J024939.57+000700.39 & 0.564 & 20.90 & 43.94 \\
DES J024646.73-001220.60 & 0.564 & 19.73 & 44.41 \\
DES J022717.88-051623.78 & 0.566 & 20.87 & 43.95 \\
DES J033946.12-295030.94 & 0.582 & 20.33 & 44.20 \\
DES J003114.43-424227.81 & 0.591 & 20.53 & 44.13 \\
DES J033905.07-292134.36 & 0.600 & 18.66 & 44.89 \\
DES J022329.27-045451.85 & 0.604 & 20.78 & 44.05 \\
DES J033738.50-272306.75 & 0.613 & 18.60 & 44.94 \\
DES J022246.23-041450.67 & 0.614 & 19.94 & 44.40 \\
DES J024651.86-010732.56 & 0.622 & 20.36 & 44.25 \\
DES J033910.13-264311.77 & 0.622 & 21.39 & 43.84 \\
DES J024442.77-004223.14 & 0.628 & 19.92 & 44.43 \\
DES J021923.29-045148.69 & 0.630 & 19.36 & 44.66 \\
DES J033051.45-271254.90 & 0.633 & 20.24 & 44.31 \\
DES J032718.70-281857.85 & 0.634 & 19.52 & 44.60 \\
DES J002904.43-425243.04 & 0.644 & 19.54 & 44.61 \\
\end{tabular}
\label{tab:AGNpropH}
\end{table}

\begin{table}
\contcaption{H$\beta$ sample}
\begin{tabular}{llll}
DES ID & $z$ & $m_r$ & $\log({\lambda}L_{5100})$ (erg\,s$^{-1}$)  \\ \hline
DES J021820.49-050426.42 & 0.650 & 19.59 & 44.60 \\
DES J034056.37-293339.67 & 0.652 & 20.44 & 44.26 \\
DES J003013.70-425733.44 $^a$ & 0.654 & 20.31 & 44.32 \\
DES J024533.65-000744.91 $^a$ & 0.655 & 19.66 & 44.58 \\
DES J003010.25-423356.22 $^a$ & 0.667 & 20.14 & 44.40 \\
DES J003231.39-433511.71 $^a$ & 0.679 & 20.39 & 44.32 \\
DES J004334.89-440546.53 $^a$ & 0.681 & 19.62 & 44.63 \\
DES J021952.14-040919.86 $^a$ & 0.692 & 20.30 & 44.37 \\
DES J022452.19-040519.38 $^a$ & 0.695 & 19.85 & 44.55 \\
DES J002906.71-423904.49 $^a$ & 0.703 & 18.69 & 45.03 \\
DES J021514.27-053321.33 $^a$ & 0.703 & 20.01 & 44.50 \\
DES J033819.34-274346.33 $^a$ & 0.706 & 18.81 & 44.98 \\
DES J022617.85-043108.99 $^a$ & 0.708 & 19.52 & 44.70 \\
DES J021524.99-045353.83 $^a$ & 0.714 & 20.51 & 44.31 \\
DES J021808.24-045845.20 $^a$ & 0.716 & 18.18 & 45.25 \\
\end{tabular}
\end{table}

\begin{table}
\caption{Properties for our OzDES Mg\,\textsc{ii} stacking sample. Columns left to right: DES name (J2000), redshift, $r$-band apparent AB magnitude, monochromatic luminosity at 3000\AA. The superscript $a$ flags sources which also have H$\beta$ data, and $b$ flags sources which also have C\,\textsc{iv} data. }
\begin{tabular}{llll}
DES ID & $z$ & $m_r$ & $\log({\lambda}L_{3000})$ (erg\,s$^{-1}$)  \\ \hline
DES J003013.70-425733.44 $^a$ & 0.654 & 20.31 & 44.47 \\
DES J024533.65-000744.91 $^a$ & 0.655 & 19.66 & 44.73 \\
DES J003010.25-423356.22 $^a$ & 0.667 & 20.14 & 44.56 \\
DES J003231.39-433511.71 $^a$ & 0.679 & 20.39 & 44.47 \\
DES J004334.89-440546.53 $^a$ & 0.681 & 19.62 & 44.78 \\
DES J021952.14-040919.86 $^a$ & 0.692 & 20.30 & 44.53 \\
DES J022452.19-040519.38 $^a$ & 0.695 & 19.85 & 44.71 \\
DES J002906.71-423904.49 $^a$ & 0.703 & 18.69 & 45.18 \\
DES J021514.27-053321.33 $^a$ & 0.703 & 20.01 & 44.66 \\
DES J033819.34-274346.33 $^a$ & 0.706 & 18.81 & 45.14 \\
DES J022617.85-043108.99 $^a$ & 0.708 & 19.52 & 44.86 \\
DES J021524.99-045353.83 $^a$ & 0.714 & 20.51 & 44.47 \\
DES J021808.24-045845.20 $^a$ & 0.716 & 18.18 & 45.40 \\
DES J033545.58-293216.49 & 0.724 & 20.30 & 44.56 \\
DES J024126.71-004526.12 & 0.727 & 18.92 & 45.12 \\
DES J033227.00-274105.28 & 0.730 & 19.81 & 44.77 \\
DES J024727.97-013008.97 & 0.741 & 20.26 & 44.60 \\
DES J022421.67-064613.71 & 0.755 & 20.48 & 44.53 \\
DES J032724.94-274202.77 & 0.756 & 19.78 & 44.81 \\
DES J021902.96-062107.22 & 0.758 & 20.42 & 44.55 \\
DES J022244.40-043346.88 & 0.761 & 19.23 & 45.03 \\
DES J025048.65+000207.62 & 0.766 & 19.12 & 45.08 \\
DES J003155.93-434225.34 & 0.768 & 19.30 & 45.01 \\
DES J021705.51-042253.46 & 0.788 & 20.24 & 44.66 \\
DES J033459.10-293317.41 & 0.801 & 19.99 & 44.77 \\
DES J032850.20-271207.84 & 0.802 & 19.42 & 45.00 \\
DES J003350.95-435606.17 & 0.806 & 19.61 & 44.93 \\
DES J022155.25-064916.59 & 0.807 & 18.06 & 45.55 \\
DES J024801.09-004015.63 & 0.811 & 20.05 & 44.76 \\
DES J025146.78-004035.49 & 0.813 & 19.63 & 44.93 \\
DES J022326.46-045705.99 & 0.825 & 20.72 & 44.50 \\
DES J025100.99+004802.99 & 0.829 & 20.03 & 44.78 \\
DES J021628.43-040147.11 & 0.831 & 19.52 & 44.99 \\
DES J033246.02-282232.12 & 0.839 & 20.07 & 44.78 \\
DES J033328.93-275641.20 & 0.839 & 20.29 & 44.69 \\
DES J022255.89-051351.61 & 0.849 & 19.01 & 45.21 \\
DES J004042.70-440341.05 & 0.851 & 19.78 & 44.91 \\
DES J033332.81-282220.34 & 0.858 & 19.14 & 45.17 \\
DES J024637.94-004105.24 & 0.859 & 20.10 & 44.79 \\
DES J033137.70-284808.03 & 0.862 & 21.68 & 44.16 \\
DES J033230.63-284750.36 & 0.862 & 20.70 & 44.55 \\
\end{tabular}
\label{tab:AGNpropM}
\end{table}

\begin{table}
\contcaption{Mg\,\textsc{ii} sample}
\begin{tabular}{llll}
DES ID & $z$ & $m_r$ & $\log({\lambda}L_{3000})$ (erg\,s$^{-1}$)  \\ \hline
DES J033235.64-290202.05 & 0.866 & 19.33 & 45.10 \\
DES J003527.80-443411.36 & 0.869 & 19.90 & 44.88 \\
DES J002831.10-421538.06 & 0.871 & 18.98 & 45.25 \\
DES J033523.52-280723.67 & 0.871 & 17.25 & 45.94 \\
DES J003331.34-441039.28 & 0.878 & 19.64 & 44.99 \\
DES J004020.36-432053.98 & 0.880 & 19.49 & 45.05 \\
DES J002641.26-424900.31 & 0.882 & 19.95 & 44.87 \\
DES J021557.63-045009.52 & 0.884 & 18.73 & 45.36 \\
DES J003301.72-440750.87 & 0.886 & 19.86 & 44.91 \\
DES J024831.08-005025.58 & 0.887 & 20.42 & 44.69 \\
DES J002951.05-433629.34 & 0.892 & 20.66 & 44.60 \\
DES J002702.17-431755.99 & 0.900 & 20.15 & 44.81 \\
DES J021413.01-042930.02 & 0.901 & 19.08 & 45.24 \\
DES J024159.74-010512.38 & 0.904 & 20.16 & 44.81 \\
DES J024357.90-011330.42 & 0.905 & 20.22 & 44.79 \\
DES J022440.72-043657.71 & 0.907 & 20.14 & 44.82 \\
DES J025217.48-005249.30 & 0.912 & 20.52 & 44.67 \\
DES J002930.76-432724.34 & 0.914 & 18.35 & 45.54 \\
DES J003437.73-424318.36 & 0.915 & 19.09 & 45.25 \\
DES J033520.08-284258.52 & 0.931 & 20.14 & 44.84 \\
DES J003435.26-441127.90 & 0.933 & 19.40 & 45.14 \\
DES J024425.38-004652.99 & 0.937 & 18.90 & 45.34 \\
DES J003809.48-435241.29 & 0.937 & 19.40 & 45.14 \\
DES J033435.50-282812.24 & 0.940 & 19.88 & 44.95 \\
DES J002926.51-431250.54 & 0.944 & 18.96 & 45.33 \\
DES J002917.61-433759.92 & 0.951 & 18.58 & 45.48 \\
DES J024237.79-012354.10 & 0.957 & 19.65 & 45.06 \\
DES J033435.49-283631.56 & 0.963 & 20.09 & 44.89 \\
DES J022049.53-053731.08 & 0.974 & 20.00 & 44.94 \\
DES J022344.57-064039.01 & 0.983 & 20.83 & 44.61 \\
DES J022712.98-044636.25 & 0.983 & 18.13 & 45.69 \\
DES J033945.79-275333.88 & 0.984 & 21.29 & 44.43 \\
DES J033404.10-275629.92 & 0.984 & 20.75 & 44.65 \\
DES J033339.36-281724.26 & 0.987 & 20.02 & 44.94 \\
DES J033509.98-283255.27 & 0.993 & 18.03 & 45.74 \\
DES J002933.85-435240.66 & 0.995 & 18.06 & 45.73 \\
DES J022114.78-050832.95 & 0.996 & 20.60 & 44.72 \\
DES J032853.99-281706.94 & 1.000 & 20.29 & 44.84 \\
DES J024514.00-003535.43 & 1.005 & 19.41 & 45.20 \\
DES J025237.80-004627.79 & 1.008 & 19.49 & 45.17 \\
DES J021500.22-043007.46 & 1.012 & 19.76 & 45.06 \\
DES J024300.44-003030.10 & 1.018 & 20.52 & 44.76 \\
DES J022449.86-043025.80 & 1.024 & 20.44 & 44.80 \\
DES J024924.67+002536.38 & 1.025 & 19.93 & 45.00 \\
DES J033531.95-271825.19 & 1.029 & 19.14 & 45.32 \\
DES J033408.25-274337.81 & 1.029 & 20.69 & 44.70 \\
DES J003914.17-443844.22 & 1.033 & 19.62 & 45.13 \\
DES J033854.56-291001.97 & 1.049 & 20.21 & 44.91 \\
DES J032831.40-285249.79 & 1.050 & 20.47 & 44.80 \\
DES J022104.63-044239.89 & 1.052 & 20.25 & 44.89 \\
DES J002826.18-433829.09 & 1.063 & 20.90 & 44.64 \\
DES J003710.86-444048.06 & 1.067 & 18.53 & 45.59 \\
DES J033810.75-275153.11 & 1.077 & 19.45 & 45.23 \\
DES J033948.28-275438.98 & 1.077 & 17.51 & 46.01 \\
DES J021817.44-045112.40 & 1.083 & 19.25 & 45.32 \\
DES J002949.39-434806.73 & 1.088 & 19.29 & 45.31 \\
DES J025228.55+003109.14 & 1.110 & 18.81 & 45.52 \\
DES J025128.38+001144.60 & 1.117 & 20.17 & 44.98 \\
DES J033045.54-284150.29 & 1.119 & 19.57 & 45.22 \\
DES J003145.76-421747.61 & 1.129 & 19.38 & 45.30 \\
DES J003834.04-432457.69 & 1.133 & 20.29 & 44.94 \\
DES J034020.28-291750.48 & 1.135 & 18.28 & 45.75 \\
DES J003536.95-443104.40 & 1.136 & 20.06 & 45.04 \\
DES J002639.87-432307.71 & 1.139 & 19.58 & 45.23 \\
\end{tabular}
\end{table}

\begin{table}
\contcaption{Mg\,\textsc{ii} sample}
\begin{tabular}{llll}
DES ID & $z$ & $m_r$ & $\log({\lambda}L_{3000})$ (erg\,s$^{-1}$)  \\ \hline
DES J024617.07-000602.60 & 1.143 & 18.42 & 45.70 \\
DES J003452.22-434613.99 & 1.146 & 20.77 & 44.76 \\
DES J024854.80+001054.04 & 1.146 & 19.57 & 45.24 \\
DES J033836.19-295113.53 & 1.148 & 20.95 & 44.69 \\
DES J003444.53-433749.06 & 1.153 & 20.12 & 45.03 \\
DES J021631.35-043207.93 & 1.155 & 19.74 & 45.18 \\
DES J022501.68-040754.18 & 1.157 & 19.67 & 45.21 \\
DES J003332.28-430526.16 & 1.159 & 18.53 & 45.67 \\
DES J022718.54-053124.86 & 1.159 & 21.06 & 44.66 \\
DES J033350.23-284244.42 & 1.161 & 21.12 & 44.63 \\
DES J021849.86-035835.29 & 1.162 & 19.18 & 45.41 \\
DES J033612.01-290152.96 & 1.164 & 19.58 & 45.25 \\
DES J033353.02-283022.38 & 1.165 & 18.51 & 45.68 \\
DES J033429.54-293904.49 & 1.167 & 20.57 & 44.86 \\
DES J033416.89-274504.73 & 1.168 & 20.12 & 45.04 \\
DES J033719.99-262418.84 & 1.169 & 19.56 & 45.27 \\
DES J003725.23-431710.85 & 1.173 & 19.94 & 45.12 \\
DES J022556.83-045853.05 & 1.180 & 21.09 & 44.66 \\
DES J022823.19-041223.67 & 1.181 & 17.57 & 46.07 \\
DES J022533.79-050801.95 & 1.183 & 18.40 & 45.74 \\
DES J033744.23-262559.74 & 1.187 & 18.85 & 45.56 \\
DES J025036.99-002408.04 & 1.195 & 20.68 & 44.84 \\
DES J033335.92-264915.24 & 1.199 & 19.99 & 45.12 \\
DES J024840.98-001228.80 & 1.199 & 19.11 & 45.47 \\
DES J003848.77-434518.55 & 1.200 & 21.03 & 44.70 \\
DES J003549.87-424526.29 & 1.208 & 20.26 & 45.02 \\
DES J033627.43-294149.95 & 1.218 & 18.59 & 45.69 \\
DES J022212.40-061246.20 & 1.220 & 20.71 & 44.85 \\
DES J022108.60-061753.21 & 1.225 & 20.00 & 45.13 \\
DES J033534.69-283149.26 & 1.225 & 19.71 & 45.25 \\
DES J022134.35-062941.64 & 1.228 & 20.71 & 44.85 \\
DES J024544.78-004415.24 & 1.229 & 19.82 & 45.21 \\
DES J022423.52-065001.40 & 1.229 & 20.15 & 45.08 \\
DES J022023.48-064959.19 & 1.232 & 20.35 & 45.00 \\
DES J033525.23-270200.72 & 1.233 & 19.31 & 45.42 \\
DES J022055.16-060136.43 & 1.234 & 20.55 & 44.92 \\
DES J003703.65-434759.50 & 1.236 & 19.87 & 45.20 \\
DES J004111.46-441014.38 & 1.237 & 20.20 & 45.06 \\
DES J033211.42-284323.98 & 1.237 & 21.86 & 44.40 \\
DES J033928.30-291714.73 & 1.241 & 20.08 & 45.12 \\
DES J003125.94-434743.50 & 1.243 & 20.25 & 45.05 \\
DES J024918.24-001731.03 & 1.244 & 19.82 & 45.22 \\
DES J021906.15-063000.89 & 1.250 & 20.12 & 45.10 \\
DES J002743.35-425258.79 & 1.252 & 19.94 & 45.18 \\
DES J033635.40-273427.02 & 1.254 & 20.17 & 45.09 \\
DES J003322.53-442412.38 & 1.257 & 19.47 & 45.37 \\
DES J022445.39-065556.80 & 1.261 & 21.11 & 44.72 \\
DES J022115.87-062217.45 & 1.273 & 18.82 & 45.64 \\
DES J024306.66-002531.29 & 1.279 & 20.36 & 45.03 \\
DES J022245.85-041932.10 & 1.283 & 20.61 & 44.93 \\
DES J034106.19-291410.86 & 1.287 & 21.20 & 44.70 \\
DES J022529.40-050946.39 & 1.288 & 20.93 & 44.81 \\
DES J024939.57-001157.73 & 1.295 & 20.97 & 44.80 \\
DES J024621.09-000152.02 & 1.296 & 18.64 & 45.73 \\
DES J021957.86-060534.66 & 1.296 & 20.74 & 44.89 \\
DES J025323.64+000446.97 & 1.297 & 20.18 & 45.12 \\
DES J022024.34-061401.95 & 1.297 & 21.31 & 44.66 \\
DES J033911.64-290601.76 & 1.298 & 20.06 & 45.16 \\
DES J025224.97+001308.30 & 1.300 & 20.68 & 44.92 \\
DES J003819.73-443134.02 & 1.305 & 18.80 & 45.67 \\
DES J024455.17-002501.41 & 1.308 & 19.29 & 45.48 \\
DES J034032.74-270521.33 & 1.310 & 20.91 & 44.83 \\
DES J024811.51-012609.51 & 1.313 & 19.29 & 45.48 \\
DES J003639.49-430400.30 & 1.317 & 20.20 & 45.12 \\
\end{tabular}
\end{table}

\begin{table}
\contcaption{Mg\,\textsc{ii} sample}
\begin{tabular}{llll}
DES ID & $z$ & $m_r$ & $\log({\lambda}L_{3000})$ (erg\,s$^{-1}$)  \\ \hline
DES J033308.48-285832.12 & 1.318 & 19.72 & 45.31 \\
DES J025005.69-004054.86 & 1.320 & 20.41 & 45.04 \\
DES J033030.59-282135.43 & 1.322 & 18.89 & 45.65 \\
DES J022024.93-053744.40 & 1.322 & 21.14 & 44.75 \\
DES J003245.16-440451.51 & 1.323 & 20.84 & 44.87 \\
DES J033216.20-273930.61 & 1.324 & 20.60 & 44.96 \\
DES J033525.44-265531.09 & 1.325 & 17.55 & 46.19 \\
DES J033939.33-272454.08 & 1.325 & 19.19 & 45.53 \\
DES J033303.99-271531.43 & 1.329 & 20.13 & 45.16 \\
DES J033533.91-264847.02 & 1.330 & 20.25 & 45.11 \\
DES J002855.47-434210.06 & 1.331 & 20.14 & 45.15 \\
DES J024214.97-003131.71 & 1.332 & 19.93 & 45.24 \\
DES J033057.32-284737.42 & 1.335 & 20.99 & 44.81 \\
DES J033635.49-263735.76 & 1.336 & 19.99 & 45.22 \\
DES J033824.55-263527.10 & 1.340 & 19.65 & 45.35 \\
DES J024652.16-011242.95 & 1.340 & 19.44 & 45.44 \\
DES J024326.61-002056.06 & 1.345 & 20.51 & 45.01 \\
DES J022351.09-053750.25 & 1.347 & 18.61 & 45.77 \\
DES J033723.98-294917.22 & 1.351 & 19.00 & 45.62 \\
DES J022446.16-050827.57 & 1.357 & 18.85 & 45.69 \\
DES J022436.17-065912.26 & 1.360 & 20.78 & 44.91 \\
DES J022419.74-062142.31 & 1.361 & 18.52 & 45.82 \\
DES J025324.69-002655.67 & 1.364 & 19.41 & 45.47 \\
DES J003922.97-430230.45 & 1.369 & 19.71 & 45.35 \\
DES J024234.93-010351.83 & 1.371 & 19.46 & 45.45 \\
DES J025252.02-002211.61 & 1.371 & 20.68 & 44.96 \\
DES J025318.92+001559.61 & 1.374 & 20.92 & 44.87 \\
DES J002857.80-424644.03 & 1.377 & 18.74 & 45.74 \\
DES J033923.53-265229.48 & 1.379 & 20.57 & 45.01 \\
DES J034027.16-285641.30 & 1.379 & 18.74 & 45.74 \\
DES J004055.63-441249.41 & 1.383 & 20.42 & 45.08 \\
DES J033444.13-264215.40 & 1.384 & 19.57 & 45.42 \\
DES J003738.22-443838.25 & 1.385 & 20.13 & 45.19 \\
DES J025145.09-004639.64 & 1.386 & 19.81 & 45.32 \\
DES J021749.09-041215.57 & 1.386 & 20.13 & 45.19 \\
DES J022815.08-041942.59 & 1.387 & 20.10 & 45.21 \\
DES J024746.99-011334.42 & 1.389 & 20.63 & 44.99 \\
DES J003814.08-433314.92 & 1.390 & 18.27 & 45.94 \\
DES J022032.25-044217.60 & 1.401 & 20.25 & 45.16 \\
DES J021451.07-044236.60 & 1.401 & 20.97 & 44.87 \\
DES J022446.95-055739.13 & 1.409 & 20.88 & 44.91 \\
DES J022402.07-062943.14 & 1.412 & 20.17 & 45.20 \\
DES J022124.53-050205.23 & 1.415 & 19.02 & 45.66 \\
DES J032942.54-272012.09 & 1.415 & 19.94 & 45.29 \\
DES J003827.69-433518.19 & 1.419 & 19.96 & 45.28 \\
DES J022436.64-063255.90 & 1.423 & 20.95 & 44.89 \\
DES J024935.55-001336.76 & 1.423 & 19.99 & 45.28 \\
DES J024606.20-005531.75 & 1.426 & 20.38 & 45.12 \\
DES J003052.76-430301.10 & 1.428 & 19.39 & 45.52 \\
DES J024929.19-002104.15 & 1.428 & 19.00 & 45.67 \\
DES J022216.72-041719.73 & 1.432 & 19.89 & 45.32 \\
DES J033918.13-293008.14 & 1.436 & 21.12 & 44.83 \\
DES J024340.09-001749.37 & 1.436 & 19.68 & 45.41 \\
DES J024753.20-002137.75 & 1.437 & 20.16 & 45.22 \\
DES J024212.65-010339.46 & 1.438 & 20.77 & 44.97 \\
DES J033435.10-262635.67 & 1.441 & 19.45 & 45.50 \\
DES J021939.92-061407.00 & 1.446 & 20.60 & 45.05 \\
DES J022041.19-055039.80 & 1.454 & 20.27 & 45.19 \\
DES J022006.60-061936.36 & 1.460 & 19.83 & 45.37 \\
DES J033630.87-291753.39 & 1.462 & 19.63 & 45.45 \\
DES J024820.90-002546.67 & 1.463 & 19.98 & 45.31 \\
DES J025030.77-000801.72 & 1.466 & 18.31 & 45.98 \\
DES J002940.83-424308.05 & 1.472 & 19.66 & 45.44 \\
DES J025024.52-001419.03 & 1.472 & 19.98 & 45.31 \\
\end{tabular}
\end{table}

\begin{table}
\contcaption{Mg\,\textsc{ii} sample}
\begin{tabular}{llll}
DES ID & $z$ & $m_r$ & $\log({\lambda}L_{3000})$ (erg\,s$^{-1}$)  \\ \hline
DES J033310.16-282433.18 & 1.476 & 20.99 & 44.91 \\
DES J003253.01-435626.33 & 1.476 & 20.01 & 45.30 \\
DES J022644.23-040720.13 & 1.479 & 19.76 & 45.41 \\
DES J003035.42-433249.60 & 1.480 & 18.77 & 45.80 \\
DES J024944.09+003317.50 & 1.480 & 19.71 & 45.43 \\
DES J003053.45-432344.99 & 1.490 & 20.58 & 45.09 \\
DES J033213.36-283621.03 & 1.492 & 20.36 & 45.18 \\
DES J003232.61-433303.00 & 1.492 & 19.77 & 45.41 \\
DES J032604.02-275629.36 & 1.494 & 18.53 & 45.91 \\
DES J022429.11-045807.69 & 1.498 & 20.30 & 45.20 \\
DES J003621.41-435139.13 & 1.501 & 20.21 & 45.24 \\
DES J034101.59-293056.44 & 1.504 & 20.49 & 45.13 \\
DES J022350.77-043158.10 & 1.504 & 19.90 & 45.37 \\
DES J021620.91-053417.45 & 1.508 & 20.11 & 45.29 \\
DES J033253.86-283139.33 & 1.509 & 20.61 & 45.09 \\
DES J024959.77-000104.11 & 1.512 & 20.09 & 45.30 \\
DES J024920.97+004206.39 & 1.524 & 20.66 & 45.08 \\
DES J021630.87-042051.45 & 1.528 & 19.90 & 45.39 \\
DES J024455.45-011500.42 & 1.529 & 20.51 & 45.15 \\
DES J033951.49-291713.89 & 1.530 & 20.83 & 45.02 \\
DES J032643.99-280657.07 & 1.532 & 19.73 & 45.46 \\
DES J004232.48-440757.18 & 1.532 & 18.38 & 46.00 \\
DES J003207.44-433049.00 & 1.533 & 19.65 & 45.50 \\
DES J033400.71-271540.89 & 1.538 & 20.04 & 45.35 \\
DES J003254.85-420236.96 & 1.541 & 20.28 & 45.25 \\
DES J022338.53-063246.90 & 1.546 & 21.24 & 44.88 \\
DES J025311.70-004241.62 & 1.547 & 20.36 & 45.23 \\
DES J022059.48-044917.03 & 1.547 & 18.51 & 45.97 \\
DES J033355.67-282651.58 & 1.551 & 20.39 & 45.22 \\
DES J033853.03-271735.34 & 1.553 & 21.16 & 44.91 \\
DES J022509.54-040838.19 & 1.553 & 20.43 & 45.21 \\
DES J003006.53-435107.85 & 1.555 & 20.46 & 45.20 \\
DES J022520.75-041246.56 & 1.559 & 20.38 & 45.23 \\
DES J022154.21-061941.56 & 1.559 & 20.13 & 45.33 \\
DES J021612.83-044634.12 & 1.560 & 19.86 & 45.44 \\
DES J003433.01-423342.34 & 1.561 & 20.28 & 45.27 \\
DES J033410.83-280949.16 & 1.562 & 21.54 & 44.77 \\
DES J033058.53-275148.31 & 1.562 & 20.90 & 45.03 \\
DES J025318.76+000414.19 & 1.563 & 20.15 & 45.33 \\
DES J033101.15-275125.44 & 1.563 & 19.12 & 45.74 \\
DES J004016.88-442556.16 & 1.565 & 20.24 & 45.29 \\
DES J003829.91-434454.27 & 1.567 & 17.83 & 46.26 \\
DES J003858.70-443045.51 & 1.569 & 19.95 & 45.41 \\
DES J003653.08-435441.53 & 1.572 & 19.32 & 45.67 \\
DES J033211.64-273726.16 & 1.574 & 19.22 & 45.71 \\
DES J033325.92-290142.07 & 1.574 & 20.77 & 45.09 \\
DES J022422.63-061943.20 & 1.575 & 21.53 & 44.78 \\
DES J002939.03-431253.68 & 1.577 & 20.70 & 45.12 \\
DES J033553.51-275044.68 & 1.578 & 18.99 & 45.80 \\
DES J025148.53-004637.61 & 1.579 & 19.50 & 45.60 \\
DES J003413.73-432600.36 & 1.588 & 21.23 & 44.91 \\
DES J022733.98-042523.34 & 1.590 & 20.57 & 45.18 \\
DES J032801.84-273815.71 & 1.590 & 20.14 & 45.35 \\
DES J003952.64-442753.23 & 1.591 & 20.03 & 45.40 \\
DES J003254.79-423926.69 & 1.597 & 19.28 & 45.70 \\
DES J003307.13-424912.19 & 1.598 & 20.49 & 45.22 \\
DES J024603.67-003211.73 $^b$ & 1.601 & 18.99 & 45.82 \\
DES J033304.62-291230.05 $^b$ & 1.603 & 20.19 & 45.34 \\
DES J024823.76-010002.36 & 1.603 & 20.48 & 45.23 \\
DES J022152.62-062834.38 $^b$ & 1.604 & 20.96 & 45.03 \\
DES J033617.72-300224.76 $^b$ & 1.605 & 19.88 & 45.47 \\
DES J033604.07-292659.88 $^b$ & 1.610 & 20.07 & 45.39 \\
DES J021505.09-045855.98 & 1.610 & 19.33 & 45.69 \\
DES J033721.40-292323.42 $^b$ & 1.612 & 19.99 & 45.43 \\
\end{tabular}
\end{table}

\begin{table}
\contcaption{Mg\,\textsc{ii} sample}
\begin{tabular}{llll}
DES ID & $z$ & $m_r$ & $\log({\lambda}L_{3000})$ (erg\,s$^{-1}$)  \\ \hline
DES J033918.18-293142.92 $^b$ & 1.612 & 21.04 & 45.01 \\
DES J033407.62-291607.55 $^b$ & 1.615 & 19.59 & 45.59 \\
DES J002950.78-423801.20 $^b$ & 1.616 & 20.69 & 45.15 \\
DES J003723.59-442258.09 $^b$ & 1.618 & 19.98 & 45.44 \\
DES J033139.43-284939.85 & 1.619 & 20.93 & 45.06 \\
DES J003426.66-422807.96 & 1.622 & 21.07 & 45.00 \\
DES J022016.53-043209.15 & 1.624 & 21.22 & 44.94 \\
DES J025225.52+003405.92 & 1.624 & 19.97 & 45.44 \\
DES J021600.36-043829.30 $^b$ & 1.624 & 20.33 & 45.30 \\
DES J033220.31-280214.82 & 1.625 & 21.11 & 44.99 \\
DES J021947.13-043754.70 $^b$ & 1.626 & 20.63 & 45.18 \\
DES J033835.80-274224.52 $^b$ & 1.629 & 20.00 & 45.44 \\
DES J003501.58-425344.26 $^b$ & 1.632 & 19.98 & 45.45 \\
DES J003454.09-425716.21 $^b$ & 1.633 & 20.62 & 45.19 \\
DES J022247.88-043330.04 $^b$ & 1.635 & 20.24 & 45.34 \\
DES J003234.33-431937.83 $^b$ & 1.641 & 19.40 & 45.68 \\
DES J022224.82-062626.69 $^b$ & 1.641 & 20.16 & 45.38 \\
DES J025254.18-001119.67 $^b$ & 1.641 & 21.25 & 44.94 \\
DES J022716.52-050008.33 $^b$ & 1.642 & 20.50 & 45.24 \\
DES J022337.73-062354.60 & 1.646 & 20.91 & 45.08 \\
DES J034044.75-270720.15 & 1.647 & 21.14 & 44.99 \\
DES J024207.50-004423.41 & 1.647 & 22.32 & 44.52 \\
DES J033437.58-275826.84 $^b$ & 1.648 & 18.69 & 45.97 \\
DES J022008.73-045905.31 $^b$ & 1.649 & 20.52 & 45.24 \\
DES J033657.91-274244.39 $^b$ & 1.649 & 20.14 & 45.39 \\
DES J003015.00-430333.45 $^b$ & 1.650 & 19.97 & 45.46 \\
DES J022029.60-051022.59 $^b$ & 1.652 & 20.68 & 45.18 \\
DES J024422.20-011247.11 & 1.654 & 20.14 & 45.40 \\
DES J022228.25-060547.60 & 1.658 & 20.34 & 45.32 \\
DES J022319.01-070927.36 & 1.659 & 21.32 & 44.93 \\
DES J003143.64-425420.21 $^b$ & 1.662 & 20.03 & 45.44 \\
DES J022208.15-065550.48 & 1.662 & 21.12 & 45.01 \\
DES J003245.74-431453.18 & 1.671 & 20.78 & 45.15 \\
DES J003548.43-431444.38 $^b$ & 1.673 & 20.47 & 45.27 \\
DES J022424.16-043229.83 $^b$ & 1.674 & 20.32 & 45.34 \\
DES J025340.94+001110.20 $^b$ & 1.675 & 18.31 & 46.14 \\
DES J024906.74+000823.79 & 1.677 & 20.68 & 45.19 \\
DES J003956.32-442047.73 $^b$ & 1.680 & 20.08 & 45.44 \\
DES J021439.12-051355.34 & 1.683 & 20.96 & 45.08 \\
DES J033903.66-293326.48 $^b$ & 1.684 & 21.85 & 44.73 \\
DES J003948.24-445323.16 $^b$ & 1.684 & 19.59 & 45.63 \\
DES J025429.12-000404.45 & 1.685 & 21.98 & 44.68 \\
DES J003635.93-434636.05 $^b$ & 1.689 & 19.90 & 45.51 \\
DES J025220.53+002735.19 $^b$ & 1.690 & 19.86 & 45.53 \\
DES J033341.61-284603.55 & 1.693 & 21.02 & 45.07 \\
DES J021902.58-044628.33 $^b$ & 1.694 & 19.92 & 45.51 \\
DES J025356.07+001057.54 $^b$ & 1.695 & 19.47 & 45.69 \\
DES J003325.69-434826.13 $^b$ & 1.696 & 20.26 & 45.37 \\
DES J024527.72-010602.84 $^b$ & 1.696 & 20.05 & 45.46 \\
DES J003030.66-420243.68 & 1.696 & 20.65 & 45.22 \\
DES J033457.77-274956.61 & 1.700 & 21.58 & 44.85 \\
DES J022445.77-061149.99 $^b$ & 1.701 & 20.19 & 45.40 \\
DES J003213.12-434553.39 & 1.706 & 17.34 & 46.55 \\
DES J003256.53-441451.15 & 1.708 & 19.89 & 45.53 \\
DES J022907.98-045102.02 & 1.714 & 20.91 & 45.12 \\
DES J022014.98-062253.15 $^b$ & 1.716 & 19.91 & 45.52 \\
DES J024207.18-002818.79 & 1.716 & 19.83 & 45.56 \\
DES J003721.82-443051.40 & 1.716 & 20.33 & 45.36 \\
DES J003221.43-423908.40 & 1.719 & 21.21 & 45.01 \\
DES J024632.55-003439.14 $^b$ & 1.724 & 19.87 & 45.55 \\
DES J022134.97-070831.84 $^b$ & 1.728 & 20.70 & 45.22 \\
DES J032940.47-275143.61 $^b$ & 1.730 & 19.87 & 45.55 \\
DES J022602.20-061626.10 $^b$ & 1.736 & 19.97 & 45.51 \\
DES J003734.68-441539.48 & 1.738 & 20.15 & 45.44 \\
\end{tabular}
\end{table}

\begin{table}
\contcaption{Mg\,\textsc{ii} sample}
\begin{tabular}{llll}
DES ID & $z$ & $m_r$ & $\log({\lambda}L_{3000})$ (erg\,s$^{-1}$)  \\ \hline
DES J024611.20-003134.33 & 1.738 & 20.52 & 45.29 \\
DES J021647.25-044029.81 & 1.746 & 21.58 & 44.87 \\
DES J003206.50-425325.23 $^b$ & 1.750 & 19.84 & 45.57 \\
DES J033814.33-294213.78 $^b$ & 1.750 & 19.94 & 45.53 \\
DES J002920.73-420425.08 & 1.750 & 19.66 & 45.64 \\
DES J033453.60-264121.34 $^b$ & 1.751 & 20.14 & 45.45 \\
DES J033817.74-290324.90 $^b$ & 1.753 & 19.61 & 45.67 \\
DES J021732.31-052950.95 & 1.755 & 21.84 & 44.77 \\
DES J033211.80-284257.07 & 1.755 & 20.26 & 45.41 \\
DES J022115.04-043155.51 $^b$ & 1.757 & 21.33 & 44.98 \\
DES J033452.89-265215.54 $^b$ & 1.757 & 19.10 & 45.87 \\
DES J033451.51-280041.23 & 1.758 & 21.49 & 44.92 \\
DES J022117.44-063856.45 $^b$ & 1.763 & 19.83 & 45.58 \\
DES J024140.99-012712.41 & 1.768 & 20.33 & 45.39 \\
DES J024257.19-004549.41 $^b$ & 1.770 & 21.08 & 45.09 \\
DES J022119.91-045948.62 $^b$ & 1.770 & 20.76 & 45.22 \\
DES J003120.86-425145.98 $^b$ & 1.771 & 20.14 & 45.47 \\
DES J002859.58-423239.47 & 1.776 & 19.43 & 45.75 \\
DES J022024.08-045709.48 & 1.776 & 19.84 & 45.59 \\
DES J033912.33-293043.57 & 1.778 & 21.35 & 44.98 \\
DES J004029.04-441704.97 $^b$ & 1.783 & 20.04 & 45.51 \\
DES J003255.12-435757.56 $^b$ & 1.784 & 20.07 & 45.50 \\
DES J022533.36-065917.56 & 1.784 & 18.95 & 45.95 \\
DES J032759.97-271921.90 $^b$ & 1.785 & 20.37 & 45.38 \\
DES J024222.95-011527.57 $^b$ & 1.785 & 20.24 & 45.43 \\
DES J032914.81-290130.69 & 1.787 & 20.50 & 45.33 \\
DES J033328.60-290314.66 $^b$ & 1.787 & 19.58 & 45.70 \\
DES J022027.32-045115.53 $^b$ & 1.788 & 21.14 & 45.08 \\
DES J033922.52-292857.78 & 1.789 & 21.27 & 45.02 \\
DES J022002.08-050101.91 $^b$ & 1.791 & 19.69 & 45.66 \\
DES J022751.50-044252.66 $^b$ & 1.795 & 20.60 & 45.30 \\
DES J003939.41-444612.90 $^b$ & 1.797 & 20.17 & 45.47 \\
DES J022503.11-065258.79 $^b$ & 1.798 & 20.09 & 45.50 \\
DES J032835.62-274406.09 $^b$ & 1.800 & 19.94 & 45.56 \\
DES J033722.89-274020.85 $^b$ & 1.804 & 19.93 & 45.57 \\
DES J003921.57-441844.65 & 1.813 & 19.91 & 45.58 \\
DES J003920.50-440441.16 $^b$ & 1.814 & 19.79 & 45.63 \\
DES J003739.59-432521.40 $^b$ & 1.815 & 20.40 & 45.39 \\
DES J022231.63-043256.29 & 1.816 & 19.67 & 45.68 \\
DES J033047.84-281521.39 & 1.818 & 21.53 & 44.94 \\
DES J034001.87-270136.24 & 1.819 & 21.05 & 45.13 \\
DES J004156.96-435856.50 & 1.821 & 20.36 & 45.41 \\
DES J024635.61-000850.46 $^b$ & 1.822 & 19.26 & 45.85 \\
DES J033002.51-274858.28 $^b$ & 1.823 & 20.70 & 45.27 \\
DES J033017.79-280231.25 & 1.823 & 20.61 & 45.31 \\
DES J033548.18-290943.65 $^b$ & 1.826 & 20.33 & 45.42 \\
DES J022352.09-064029.96 $^b$ & 1.827 & 20.50 & 45.35 \\
DES J022409.89-044718.00 $^b$ & 1.832 & 20.11 & 45.51 \\
DES J033319.58-290431.65 $^b$ & 1.833 & 20.73 & 45.27 \\
DES J024716.32-010401.36 $^b$ & 1.834 & 21.10 & 45.12 \\
DES J021958.13-041707.68 & 1.837 & 21.09 & 45.12 \\
DES J033222.47-285935.09 $^b$ & 1.837 & 20.68 & 45.29 \\
DES J033301.69-275818.80 $^b$ & 1.840 & 20.94 & 45.19 \\
DES J004014.18-431716.29 $^b$ & 1.842 & 19.87 & 45.62 \\
DES J021643.70-052236.21 & 1.844 & 20.57 & 45.34 \\
DES J021903.49-043935.00 $^b$ & 1.845 & 20.48 & 45.37 \\
DES J033539.94-265025.37 & 1.846 & 19.68 & 45.69 \\
DES J024737.50-000458.97 $^b$ & 1.847 & 19.88 & 45.62 \\
DES J024608.67-013933.97 $^b$ & 1.848 & 19.68 & 45.70 \\
DES J022515.35-044008.91 $^b$ & 1.850 & 19.82 & 45.64 \\
DES J025038.68-004739.06 $^b$ & 1.856 & 18.67 & 46.10 \\
DES J002729.28-431501.02 $^b$ & 1.863 & 19.67 & 45.71 \\
DES J024723.54-002536.48 $^b$ & 1.864 & 19.42 & 45.81 \\
DES J022845.57-043350.18 $^b$ & 1.864 & 17.41 & 46.61 \\
\end{tabular}
\end{table}

\begin{table}
\contcaption{Mg\,\textsc{ii} sample}
\begin{tabular}{llll}
DES ID & $z$ & $m_r$ & $\log({\lambda}L_{3000})$ (erg\,s$^{-1}$)  \\ \hline
DES J024748.87-000147.41 $^b$ & 1.868 & 19.58 & 45.75 \\
DES J024133.65-010724.20 $^b$ & 1.874 & 20.86 & 45.24 \\
DES J003940.80-443718.32 $^b$ & 1.875 & 19.49 & 45.79 \\
DES J024838.93-000325.87 & 1.876 & 20.61 & 45.34 \\
DES J022047.63-051835.83 $^b$ & 1.878 & 19.55 & 45.77 \\
DES J021544.02-053607.56 & 1.885 & 21.31 & 45.06 \\
DES J025513.03+000639.59 & 1.885 & 19.78 & 45.68 \\
DES J021707.96-055152.07 $^b$ & 1.885 & 19.70 & 45.71 \\
DES J033145.20-275435.75 $^b$ & 1.891 & 21.19 & 45.12 \\
DES J033225.90-282208.45 & 1.894 & 21.55 & 44.97 \\
DES J022213.31-041030.05 & 1.903 & 20.14 & 45.54 \\
DES J024242.50-002212.87 $^b$ & 1.904 & 20.57 & 45.37 \\
DES J022828.19-040044.25 $^b$ & 1.905 & 18.31 & 46.28 \\
DES J003626.44-430820.66 $^b$ & 1.907 & 19.80 & 45.68 \\
DES J022050.39-061748.80 $^b$ & 1.908 & 21.89 & 44.85 \\
DES J021953.45-061123.39 $^b$ & 1.910 & 19.90 & 45.64 \\
DES J003150.13-432726.88 $^b$ & 1.912 & 18.68 & 46.13 \\
DES J022328.88-040134.71 $^b$ & 1.915 & 19.82 & 45.68 \\
DES J033739.88-272045.97 $^b$ & 1.916 & 19.21 & 45.92 \\
DES J033924.44-271250.15 $^b$ & 1.918 & 20.52 & 45.40 \\
DES J025428.31+002418.43 $^b$ & 1.918 & 20.42 & 45.44 \\
DES J033415.37-265535.62 $^b$ & 1.922 & 19.85 & 45.67 \\
DES J033011.77-283636.65 $^b$ & 1.922 & 20.99 & 45.21 \\
DES J022327.85-040119.16 $^b$ & 1.922 & 20.81 & 45.29 \\
DES J024639.22-012731.85 $^b$ & 1.923 & 20.34 & 45.47 \\
\end{tabular}
\end{table}

\begin{table}
\caption{Properties for our OzDES C\,\textsc{iv} stacking sample. Columns left to right: DES name (J2000), redshift, $r$-band apparent AB magnitude, monochromatic luminosity at 1350\AA. The superscript $b$ flags sources which also have Mg\,\textsc{ii} data. }
\begin{tabular}{llll}
DES ID & $z$ & $m_r$ & $\log({\lambda}L_{1350})$ (erg\,s$^{-1}$)  \\ \hline
DES J024603.67-003211.73 $^a$ & 1.601 & 18.99 & 45.93 \\
DES J033304.62-291230.05 $^a$ & 1.603 & 20.19 & 45.45 \\
DES J022152.62-062834.38 $^a$ & 1.604 & 20.96 & 45.14 \\
DES J033617.72-300224.76 $^a$ & 1.605 & 19.88 & 45.57 \\
DES J033604.07-292659.88 $^a$ & 1.610 & 20.07 & 45.50 \\
DES J033721.40-292323.42 $^a$ & 1.612 & 19.99 & 45.53 \\
DES J033918.18-293142.92 $^a$ & 1.612 & 21.04 & 45.11 \\
DES J033407.62-291607.55 $^a$ & 1.615 & 19.59 & 45.70 \\
DES J002950.78-423801.20 $^a$ & 1.616 & 20.69 & 45.26 \\
DES J003723.59-442258.09 $^a$ & 1.618 & 19.98 & 45.54 \\
DES J021600.36-043829.30 $^a$ & 1.624 & 20.33 & 45.41 \\
DES J021947.13-043754.70 $^a$ & 1.626 & 20.63 & 45.29 \\
DES J033835.80-274224.52 $^a$ & 1.629 & 20.00 & 45.54 \\
DES J003501.58-425344.26 $^a$ & 1.632 & 19.98 & 45.55 \\
DES J003454.09-425716.21 $^a$ & 1.633 & 20.62 & 45.30 \\
DES J022247.88-043330.04 $^a$ & 1.635 & 20.24 & 45.45 \\
DES J003234.33-431937.83 $^a$ & 1.641 & 19.40 & 45.79 \\
DES J022224.82-062626.69 $^a$ & 1.641 & 20.16 & 45.48 \\
DES J025254.18-001119.67 $^a$ & 1.641 & 21.25 & 45.05 \\
DES J022716.52-050008.33 $^a$ & 1.642 & 20.50 & 45.35 \\
DES J033437.58-275826.84 $^a$ & 1.648 & 18.69 & 46.08 \\
DES J022008.73-045905.31 $^a$ & 1.649 & 20.52 & 45.35 \\
DES J033657.91-274244.39 $^a$ & 1.649 & 20.14 & 45.50 \\
DES J003015.00-430333.45 $^a$ & 1.650 & 19.97 & 45.57 \\
DES J022029.60-051022.59 $^a$ & 1.652 & 20.68 & 45.28 \\
DES J003143.64-425420.21 $^a$ & 1.662 & 20.03 & 45.55 \\
DES J003548.43-431444.38 $^a$ & 1.673 & 20.47 & 45.38 \\
DES J022424.16-043229.83 $^a$ & 1.674 & 20.32 & 45.44 \\
DES J025340.94+001110.20 $^a$ & 1.675 & 18.31 & 46.25 \\
DES J003956.32-442047.73 $^a$ & 1.680 & 20.08 & 45.54 \\
DES J033903.66-293326.48 $^a$ & 1.684 & 21.85 & 44.84 \\
\end{tabular}
\label{tab:AGNpropC}
\end{table}

\begin{table}
\contcaption{C\,\textsc{iv} sample}
\begin{tabular}{llll}
DES ID & $z$ & $m_r$ & $\log({\lambda}L_{1350})$ (erg\,s$^{-1}$)  \\ \hline
DES J003948.24-445323.16 $^a$ & 1.684 & 19.59 & 45.74 \\
DES J003635.93-434636.05 $^a$ & 1.689 & 19.90 & 45.62 \\
DES J025220.53+002735.19 $^a$ & 1.690 & 19.86 & 45.63 \\
DES J021902.58-044628.33 $^a$ & 1.694 & 19.92 & 45.61 \\
DES J025356.07+001057.54 $^a$ & 1.695 & 19.47 & 45.79 \\
DES J003325.69-434826.13 $^a$ & 1.696 & 20.26 & 45.48 \\
DES J024527.72-010602.84 $^a$ & 1.696 & 20.05 & 45.56 \\
DES J022445.77-061149.99 $^a$ & 1.701 & 20.19 & 45.51 \\
DES J022014.98-062253.15 $^a$ & 1.716 & 19.91 & 45.63 \\
DES J024632.55-003439.14 $^a$ & 1.724 & 19.87 & 45.65 \\
DES J022134.97-070831.84 $^a$ & 1.728 & 20.70 & 45.32 \\
DES J032940.47-275143.61 $^a$ & 1.730 & 19.87 & 45.66 \\
DES J022602.20-061626.10 $^a$ & 1.736 & 19.97 & 45.62 \\
DES J003206.50-425325.23 $^a$ & 1.750 & 19.84 & 45.68 \\
DES J033814.33-294213.78 $^a$ & 1.750 & 19.94 & 45.64 \\
DES J033453.60-264121.34 $^a$ & 1.751 & 20.14 & 45.56 \\
DES J033817.74-290324.90 $^a$ & 1.753 & 19.61 & 45.77 \\
DES J022115.04-043155.51 $^a$ & 1.757 & 21.33 & 45.09 \\
DES J033452.89-265215.54 $^a$ & 1.757 & 19.10 & 45.98 \\
DES J022117.44-063856.45 $^a$ & 1.763 & 19.83 & 45.69 \\
DES J024257.19-004549.41 $^a$ & 1.770 & 21.08 & 45.20 \\
DES J022119.91-045948.62 $^a$ & 1.770 & 20.76 & 45.32 \\
DES J003120.86-425145.98 $^a$ & 1.771 & 20.14 & 45.57 \\
DES J004029.04-441704.97 $^a$ & 1.783 & 20.04 & 45.62 \\
DES J003255.12-435757.56 $^a$ & 1.784 & 20.07 & 45.61 \\
DES J032759.97-271921.90 $^a$ & 1.785 & 20.37 & 45.49 \\
DES J024222.95-011527.57 $^a$ & 1.785 & 20.24 & 45.54 \\
DES J033328.60-290314.66 $^a$ & 1.787 & 19.58 & 45.80 \\
DES J022027.32-045115.53 $^a$ & 1.788 & 21.14 & 45.18 \\
DES J022002.08-050101.91 $^a$ & 1.791 & 19.69 & 45.76 \\
DES J022751.50-044252.66 $^a$ & 1.795 & 20.60 & 45.40 \\
DES J003939.41-444612.90 $^a$ & 1.797 & 20.17 & 45.58 \\
DES J022503.11-065258.79 $^a$ & 1.798 & 20.09 & 45.61 \\
DES J032835.62-274406.09 $^a$ & 1.800 & 19.94 & 45.67 \\
DES J033722.89-274020.85 $^a$ & 1.804 & 19.93 & 45.67 \\
DES J003920.50-440441.16 $^a$ & 1.814 & 19.79 & 45.74 \\
DES J003739.59-432521.40 $^a$ & 1.815 & 20.40 & 45.49 \\
DES J024635.61-000850.46 $^a$ & 1.822 & 19.26 & 45.95 \\
DES J033002.51-274858.28 $^a$ & 1.823 & 20.70 & 45.38 \\
DES J033548.18-290943.65 $^a$ & 1.826 & 20.33 & 45.53 \\
DES J022352.09-064029.96 $^a$ & 1.827 & 20.50 & 45.46 \\
DES J022409.89-044718.00 $^a$ & 1.832 & 20.11 & 45.62 \\
DES J033319.58-290431.65 $^a$ & 1.833 & 20.73 & 45.37 \\
DES J024716.32-010401.36 $^a$ & 1.834 & 21.10 & 45.23 \\
DES J033222.47-285935.09 $^a$ & 1.837 & 20.68 & 45.40 \\
DES J033301.69-275818.80 $^a$ & 1.840 & 20.94 & 45.29 \\
DES J004014.18-431716.29 $^a$ & 1.842 & 19.87 & 45.72 \\
DES J021903.49-043935.00 $^a$ & 1.845 & 20.48 & 45.48 \\
DES J024737.50-000458.97 $^a$ & 1.847 & 19.88 & 45.72 \\
DES J024608.67-013933.97 $^a$ & 1.848 & 19.68 & 45.80 \\
DES J022515.35-044008.91 $^a$ & 1.850 & 19.82 & 45.75 \\
DES J025038.68-004739.06 $^a$ & 1.856 & 18.67 & 46.21 \\
DES J002729.28-431501.02 $^a$ & 1.863 & 19.67 & 45.81 \\
DES J024723.54-002536.48 $^a$ & 1.864 & 19.42 & 45.92 \\
DES J022845.57-043350.18 $^a$ & 1.864 & 17.41 & 46.72 \\
DES J024748.87-000147.41 $^a$ & 1.868 & 19.58 & 45.85 \\
DES J024133.65-010724.20 $^a$ & 1.874 & 20.86 & 45.34 \\
DES J003940.80-443718.32 $^a$ & 1.875 & 19.49 & 45.89 \\
DES J022047.63-051835.83 $^a$ & 1.878 & 19.55 & 45.87 \\
DES J021707.96-055152.07 $^a$ & 1.885 & 19.70 & 45.82 \\
DES J033145.20-275435.75 $^a$ & 1.891 & 21.19 & 45.22 \\
DES J024242.50-002212.87 $^a$ & 1.904 & 20.57 & 45.48 \\
DES J022828.19-040044.25 $^a$ & 1.905 & 18.31 & 46.38 \\
DES J003626.44-430820.66 $^a$ & 1.907 & 19.80 & 45.79 \\
\end{tabular}
\end{table}

\begin{table}
\contcaption{C\,\textsc{iv} sample}
\begin{tabular}{llll}
DES ID & $z$ & $m_r$ & $\log({\lambda}L_{1350})$ (erg\,s$^{-1}$)  \\ \hline
DES J022050.39-061748.80 $^a$ & 1.908 & 21.89 & 44.95 \\
DES J021953.45-061123.39 $^a$ & 1.910 & 19.90 & 45.75 \\
DES J003150.13-432726.88 $^a$ & 1.912 & 18.68 & 46.24 \\
DES J022328.88-040134.71 $^a$ & 1.915 & 19.82 & 45.78 \\
DES J033739.88-272045.97 $^a$ & 1.916 & 19.21 & 46.03 \\
DES J033924.44-271250.15 $^a$ & 1.918 & 20.52 & 45.51 \\
DES J025428.31+002418.43 $^a$ & 1.918 & 20.42 & 45.55 \\
DES J033415.37-265535.62 $^a$ & 1.922 & 19.85 & 45.78 \\
DES J033011.77-283636.65 $^a$ & 1.922 & 20.99 & 45.32 \\
DES J022327.85-040119.16 $^a$ & 1.922 & 20.81 & 45.39 \\
DES J024639.22-012731.85 $^a$ & 1.923 & 20.34 & 45.58 \\
DES J022250.47-060104.89 & 1.925 & 21.54 & 45.10 \\
DES J022206.47-060746.70 & 1.927 & 20.89 & 45.36 \\
DES J022514.39-044700.14 & 1.928 & 18.83 & 46.19 \\
DES J034005.69-292312.82 & 1.928 & 19.04 & 46.10 \\
DES J033701.63-282753.94 & 1.933 & 20.20 & 45.64 \\
DES J034016.06-270925.57 & 1.935 & 19.72 & 45.84 \\
DES J033832.50-291215.65 & 1.935 & 20.77 & 45.41 \\
DES J022537.03-050109.34 & 1.936 & 20.03 & 45.71 \\
DES J033316.17-292917.48 & 1.947 & 19.95 & 45.75 \\
DES J033607.36-263207.88 & 1.948 & 20.02 & 45.72 \\
DES J033401.79-265054.28 & 1.952 & 19.72 & 45.84 \\
DES J033232.00-280309.98 & 1.954 & 19.98 & 45.74 \\
DES J025233.79+004340.74 & 1.954 & 20.60 & 45.49 \\
DES J033646.40-291617.17 & 1.956 & 20.47 & 45.54 \\
DES J033052.19-274926.84 & 1.958 & 21.49 & 45.14 \\
DES J033559.60-280324.41 & 1.965 & 20.60 & 45.49 \\
DES J024918.63+000548.34 & 1.966 & 20.66 & 45.47 \\
DES J033650.66-293142.07 & 1.972 & 21.68 & 45.06 \\
DES J032926.49-271844.01 & 1.978 & 20.33 & 45.61 \\
DES J021949.99-064208.13 & 1.984 & 19.93 & 45.77 \\
DES J033839.06-292351.17 & 1.987 & 20.33 & 45.61 \\
DES J025159.70-005159.89 & 1.988 & 21.84 & 45.01 \\
DES J022812.23-043227.58 & 1.992 & 21.04 & 45.33 \\
DES J004016.18-435116.09 & 1.995 & 19.98 & 45.75 \\
DES J033912.95-273516.85 & 2.006 & 20.08 & 45.72 \\
DES J032821.07-282055.17 & 2.011 & 19.50 & 45.95 \\
DES J022352.19-043031.68 & 2.012 & 19.75 & 45.85 \\
DES J033040.28-274203.95 & 2.013 & 19.96 & 45.77 \\
DES J022657.24-035944.42 & 2.018 & 20.17 & 45.68 \\
DES J025019.26+003100.81 & 2.018 & 20.26 & 45.65 \\
DES J022530.19-065458.19 & 2.030 & 19.49 & 45.96 \\
DES J032703.62-274425.27 & 2.031 & 19.45 & 45.98 \\
DES J022034.93-052956.02 & 2.036 & 18.20 & 46.48 \\
DES J034036.10-284811.29 & 2.039 & 20.13 & 45.71 \\
DES J022629.26-043057.07 & 2.040 & 20.18 & 45.69 \\
DES J002959.21-434835.24 & 2.041 & 17.77 & 46.65 \\
DES J022045.50-045121.91 & 2.045 & 19.92 & 45.80 \\
DES J021935.78-064227.17 & 2.060 & 21.83 & 45.04 \\
DES J003146.74-423601.03 & 2.061 & 19.54 & 45.95 \\
DES J033635.40-280828.78 & 2.067 & 20.28 & 45.66 \\
DES J021905.63-055958.78 & 2.069 & 20.11 & 45.73 \\
DES J003549.05-440236.08 & 2.072 & 19.23 & 46.08 \\
DES J021957.25-043952.46 & 2.077 & 18.90 & 46.22 \\
DES J004143.04-440234.42 & 2.081 & 20.32 & 45.65 \\
DES J024453.25-011256.33 & 2.083 & 20.73 & 45.49 \\
DES J024344.30-000201.08 & 2.083 & 20.04 & 45.77 \\
DES J032602.35-281031.00 & 2.085 & 20.01 & 45.78 \\
DES J021850.07-050954.10 & 2.091 & 19.52 & 45.98 \\
DES J021921.81-043642.21 & 2.092 & 20.04 & 45.77 \\
DES J021941.16-044100.36 & 2.096 & 21.15 & 45.33 \\
DES J033015.97-280219.79 & 2.097 & 20.18 & 45.72 \\
DES J004022.52-442820.25 & 2.100 & 19.42 & 46.02 \\
DES J022431.59-052818.78 & 2.102 & 17.53 & 46.78 \\
\end{tabular}
\end{table}

\begin{table}
\contcaption{C\,\textsc{iv} sample}
\begin{tabular}{llll}
DES ID & $z$ & $m_r$ & $\log({\lambda}L_{1350})$ (erg\,s$^{-1}$)  \\ \hline
DES J024628.49-004457.13 & 2.104 & 19.94 & 45.81 \\
DES J003848.52-440202.07 & 2.105 & 19.72 & 45.90 \\
DES J024357.91-005447.75 & 2.105 & 21.03 & 45.38 \\
DES J022044.45-050905.95 & 2.108 & 21.19 & 45.32 \\
DES J025209.79+000549.21 & 2.108 & 19.95 & 45.81 \\
DES J024756.35-001555.92 & 2.119 & 18.70 & 46.32 \\
DES J022813.61-042251.65 & 2.121 & 19.63 & 45.95 \\
DES J033640.29-284423.00 & 2.124 & 20.80 & 45.48 \\
DES J024739.11-005221.01 & 2.124 & 20.50 & 45.60 \\
DES J022411.52-060915.92 & 2.128 & 19.82 & 45.87 \\
DES J024550.79-004328.13 & 2.131 & 20.52 & 45.59 \\
DES J003816.16-434647.44 & 2.141 & 19.54 & 45.99 \\
DES J003936.81-443439.72 & 2.144 & 19.97 & 45.82 \\
DES J022219.46-062539.66 & 2.146 & 19.04 & 46.19 \\
DES J032829.96-274212.23 & 2.150 & 20.12 & 45.76 \\
DES J024632.44-003214.12 & 2.152 & 18.89 & 46.25 \\
DES J021441.86-043709.03 & 2.152 & 22.17 & 44.94 \\
DES J022028.95-045802.58 & 2.153 & 20.53 & 45.60 \\
DES J033326.24-275829.85 & 2.161 & 20.53 & 45.60 \\
DES J022351.07-044729.93 & 2.163 & 20.93 & 45.44 \\
DES J033217.15-271956.41 & 2.169 & 19.24 & 46.12 \\
DES J032923.16-280022.78 & 2.175 & 19.77 & 45.91 \\
DES J002913.21-433044.56 & 2.180 & 20.11 & 45.78 \\
DES J025335.12-003039.26 & 2.184 & 20.41 & 45.66 \\
DES J032939.97-284952.40 & 2.188 & 20.22 & 45.73 \\
DES J024557.22-000823.36 & 2.199 & 20.05 & 45.81 \\
DES J022001.63-052216.92 & 2.219 & 20.17 & 45.77 \\
DES J003320.08-430428.45 & 2.231 & 20.19 & 45.76 \\
DES J024840.78-003548.15 & 2.235 & 18.89 & 46.28 \\
DES J022358.64-045351.63 & 2.235 & 20.29 & 45.72 \\
DES J003448.83-442519.26 & 2.237 & 20.13 & 45.79 \\
DES J033852.79-282256.93 & 2.246 & 19.78 & 45.93 \\
DES J022402.32-061740.34 & 2.246 & 19.91 & 45.88 \\
DES J002907.77-420916.39 & 2.247 & 20.77 & 45.54 \\
DES J024907.29-000649.35 & 2.251 & 20.31 & 45.72 \\
DES J003743.89-434715.68 & 2.257 & 19.60 & 46.01 \\
DES J022725.81-033837.75 & 2.257 & 18.20 & 46.57 \\
DES J021757.52-045059.15 & 2.257 & 20.73 & 45.56 \\
DES J003516.08-432451.82 & 2.260 & 19.47 & 46.06 \\
DES J022557.62-045005.32 & 2.271 & 19.55 & 46.03 \\
DES J022229.62-044941.93 & 2.275 & 21.22 & 45.37 \\
DES J033604.00-274203.95 & 2.277 & 21.41 & 45.29 \\
DES J022729.22-043227.66 & 2.280 & 19.75 & 45.96 \\
DES J033655.83-290218.23 & 2.283 & 19.41 & 46.09 \\
DES J024204.58-003835.64 & 2.287 & 20.21 & 45.77 \\
DES J022451.98-041210.79 & 2.294 & 20.14 & 45.80 \\
DES J022612.64-043401.32 & 2.306 & 21.08 & 45.43 \\
DES J022748.85-042820.90 & 2.311 & 20.08 & 45.84 \\
DES J022410.96-050653.95 & 2.315 & 20.66 & 45.60 \\
DES J024950.74-004224.16 & 2.317 & 20.45 & 45.69 \\
DES J024520.05-004534.45 & 2.321 & 20.79 & 45.55 \\
DES J021730.92-041823.54 & 2.322 & 19.75 & 45.97 \\
DES J033843.76-294922.54 & 2.328 & 20.67 & 45.60 \\
DES J022824.68-041545.71 & 2.328 & 19.63 & 46.02 \\
DES J003436.78-440043.41 & 2.330 & 19.58 & 46.04 \\
DES J025128.73-002650.64 & 2.330 & 19.95 & 45.89 \\
DES J022055.10-061842.21 & 2.340 & 21.09 & 45.44 \\
DES J024337.15-002340.17 & 2.347 & 19.36 & 46.14 \\
DES J033822.77-275910.73 & 2.349 & 19.64 & 46.02 \\
DES J024505.03-003441.97 & 2.350 & 20.60 & 45.64 \\
DES J025033.20+003829.22 & 2.364 & 19.96 & 45.90 \\
DES J004056.56-431446.40 & 2.384 & 19.56 & 46.07 \\
DES J033734.80-294213.92 & 2.390 & 19.84 & 45.96 \\
DES J033953.34-270053.36 & 2.410 & 18.53 & 46.49 \\
\end{tabular}
\end{table}

\begin{table}
\contcaption{C\,\textsc{iv} sample}
\begin{tabular}{llll}
DES ID & $z$ & $m_r$ & $\log({\lambda}L_{1350})$ (erg\,s$^{-1}$)  \\ \hline
DES J003430.47-433311.36 & 2.415 & 20.75 & 45.60 \\
DES J025102.06-004142.78 & 2.425 & 22.15 & 45.05 \\
DES J032942.28-281906.69 & 2.436 & 20.42 & 45.74 \\
DES J033605.57-290354.17 & 2.450 & 20.17 & 45.85 \\
DES J022354.81-044814.94 & 2.452 & 18.74 & 46.42 \\
DES J022014.33-042917.10 & 2.458 & 19.30 & 46.20 \\
DES J024511.94-011317.50 & 2.462 & 20.46 & 45.74 \\
DES J024757.67+001542.65 & 2.467 & 20.87 & 45.57 \\
DES J033430.30-275958.91 & 2.472 & 21.32 & 45.40 \\
DES J022540.58-043825.15 & 2.477 & 20.27 & 45.82 \\
DES J032640.93-283206.80 & 2.492 & 20.28 & 45.82 \\
DES J024717.40-000052.27 & 2.497 & 20.30 & 45.81 \\
DES J003017.36-423144.34 & 2.497 & 20.00 & 45.93 \\
DES J003957.42-434107.92 & 2.500 & 20.14 & 45.88 \\
DES J033237.62-271448.07 & 2.519 & 20.30 & 45.82 \\
DES J033608.10-285245.81 & 2.522 & 20.30 & 45.82 \\
DES J003145.23-424618.45 & 2.523 & 20.70 & 45.66 \\
DES J033545.52-275451.69 & 2.540 & 20.28 & 45.83 \\
DES J003341.34-420811.51 & 2.541 & 19.80 & 46.02 \\
DES J033216.69-272411.37 & 2.547 & 19.77 & 46.04 \\
DES J022259.87-063326.65 & 2.563 & 20.60 & 45.71 \\
DES J003547.38-430558.81 & 2.565 & 20.03 & 45.94 \\
DES J022434.33-043200.27 & 2.568 & 20.06 & 45.93 \\
DES J033518.30-275304.01 & 2.576 & 19.33 & 46.22 \\
DES J003530.07-444539.88 & 2.587 & 20.06 & 45.93 \\
DES J003352.72-425452.55 & 2.593 & 19.10 & 46.32 \\
DES J033331.37-275634.38 & 2.602 & 20.37 & 45.81 \\
DES J024719.59-003313.11 & 2.607 & 20.48 & 45.77 \\
DES J032739.71-281934.84 & 2.616 & 20.18 & 45.89 \\
DES J034121.52-265901.04 & 2.621 & 19.44 & 46.19 \\
DES J021457.21-043011.44 & 2.636 & 18.34 & 46.63 \\
DES J022104.88-060728.52 & 2.648 & 20.05 & 45.95 \\
DES J022631.82-045127.47 & 2.655 & 20.77 & 45.66 \\
DES J022330.15-043004.09 & 2.677 & 20.78 & 45.66 \\
DES J021719.45-052305.27 & 2.695 & 19.36 & 46.23 \\
DES J022034.52-045132.47 & 2.732 & 20.40 & 45.83 \\
DES J033944.09-284604.07 & 2.744 & 19.33 & 46.27 \\
DES J022620.86-045946.48 & 2.745 & 21.30 & 45.48 \\
DES J025223.48-003623.57 & 2.769 & 20.20 & 45.93 \\
DES J033526.38-285747.93 & 2.770 & 18.08 & 46.78 \\
DES J033938.51-291019.54 & 2.778 & 21.16 & 45.55 \\
DES J021921.22-044315.03 & 2.794 & 20.17 & 45.95 \\
DES J022636.07-043428.92 & 2.801 & 19.88 & 46.07 \\
DES J022354.85-054839.89 & 2.814 & 19.31 & 46.30 \\
DES J021659.87-053203.49 & 2.818 & 17.60 & 46.99 \\
DES J022305.97-054015.26 & 2.823 & 20.04 & 46.01 \\
DES J022321.26-055600.27 & 2.846 & 21.55 & 45.42 \\
DES J022255.51-044410.32 & 2.871 & 20.31 & 45.92 \\
DES J033720.58-272434.70 & 2.896 & 20.16 & 45.99 \\
DES J003255.24-430948.41 & 3.010 & 19.57 & 46.26 \\
DES J022423.43-070627.92 & 3.030 & 21.48 & 45.50 \\
DES J003318.29-434301.10 & 3.082 & 19.31 & 46.38 \\
DES J033246.76-280846.78 & 3.180 & 19.17 & 46.47 \\
DES J021438.15-052024.38 & 3.214 & 20.93 & 45.77 \\
DES J025250.10+000830.68 & 3.237 & 20.27 & 46.04 \\
DES J003133.50-422953.97 & 3.248 & 20.62 & 45.90 \\
DES J021906.24-041933.94 & 3.330 & 21.60 & 45.54 \\
DES J033712.16-285146.74 & 3.364 & 20.33 & 46.06 \\
DES J003411.44-424329.48 & 3.373 & 20.83 & 45.86 \\
DES J002926.44-420752.35 & 3.403 & 19.71 & 46.32 \\
DES J022543.53-042834.48 & 3.411 & 20.64 & 45.95 \\
DES J024325.96-003145.11 & 3.435 & 19.63 & 46.36 \\
DES J022133.35-065713.47 & 3.440 & 20.16 & 46.15 \\
DES J034122.60-291302.04 & 3.470 & 20.33 & 46.09 \\
\end{tabular}
\end{table}

\begin{table}
\contcaption{C\,\textsc{iv} sample}
\begin{tabular}{llll}
DES ID & $z$ & $m_r$ & $\log({\lambda}L_{1350})$ (erg\,s$^{-1}$)  \\ \hline
DES J033512.61-292351.14 & 3.712 & 19.98 & 46.27 \\
DES J024509.76-010001.28 & 3.807 & 20.58 & 46.05 \\
DES J022144.21-062745.18 & 3.856 & 20.14 & 46.24 \\
\end{tabular}
\end{table}

\section{H$\beta$ excluding objects with individual lags}\label{sec:appendixB}

\begin{figure}
    \centering
    \includegraphics[width=\columnwidth]{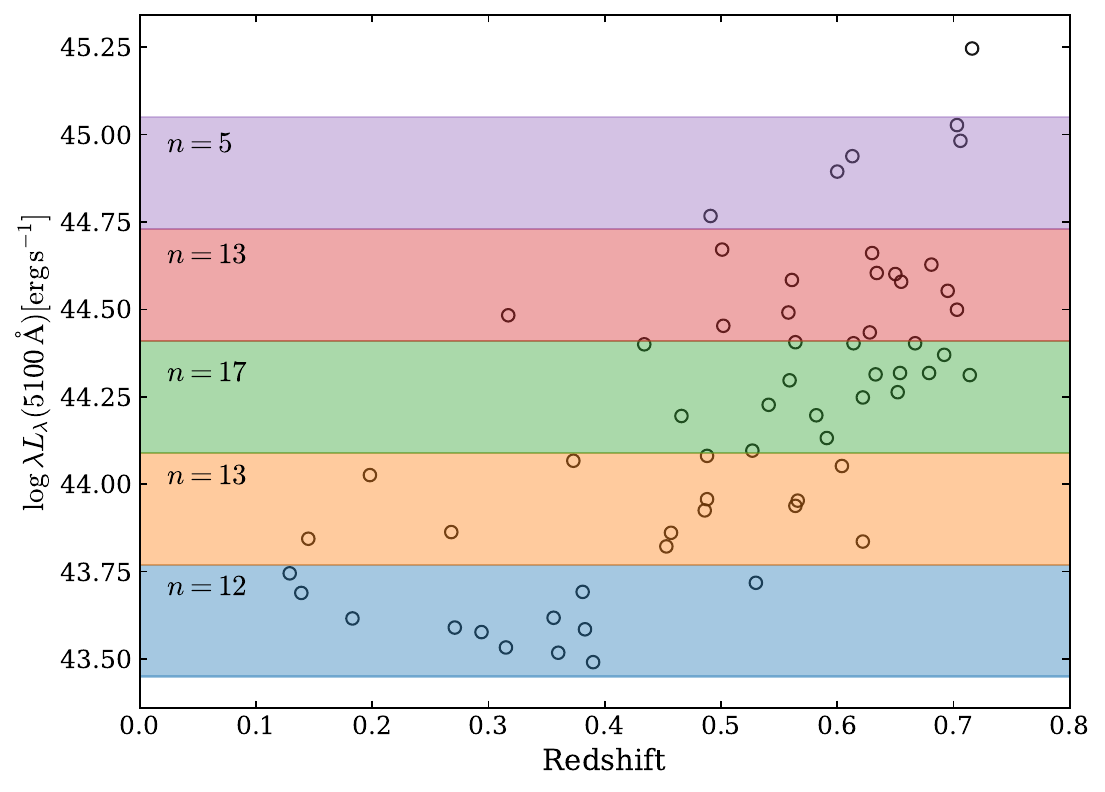}
    \caption{The same as \autoref{fig:Hbeta_bins} after excluding sources with individual lag recoveries. }
    \label{fig:Hbeta_bins_r_i}
\end{figure}

\begin{figure*}
    \centering
    \includegraphics[width=0.9\textwidth]{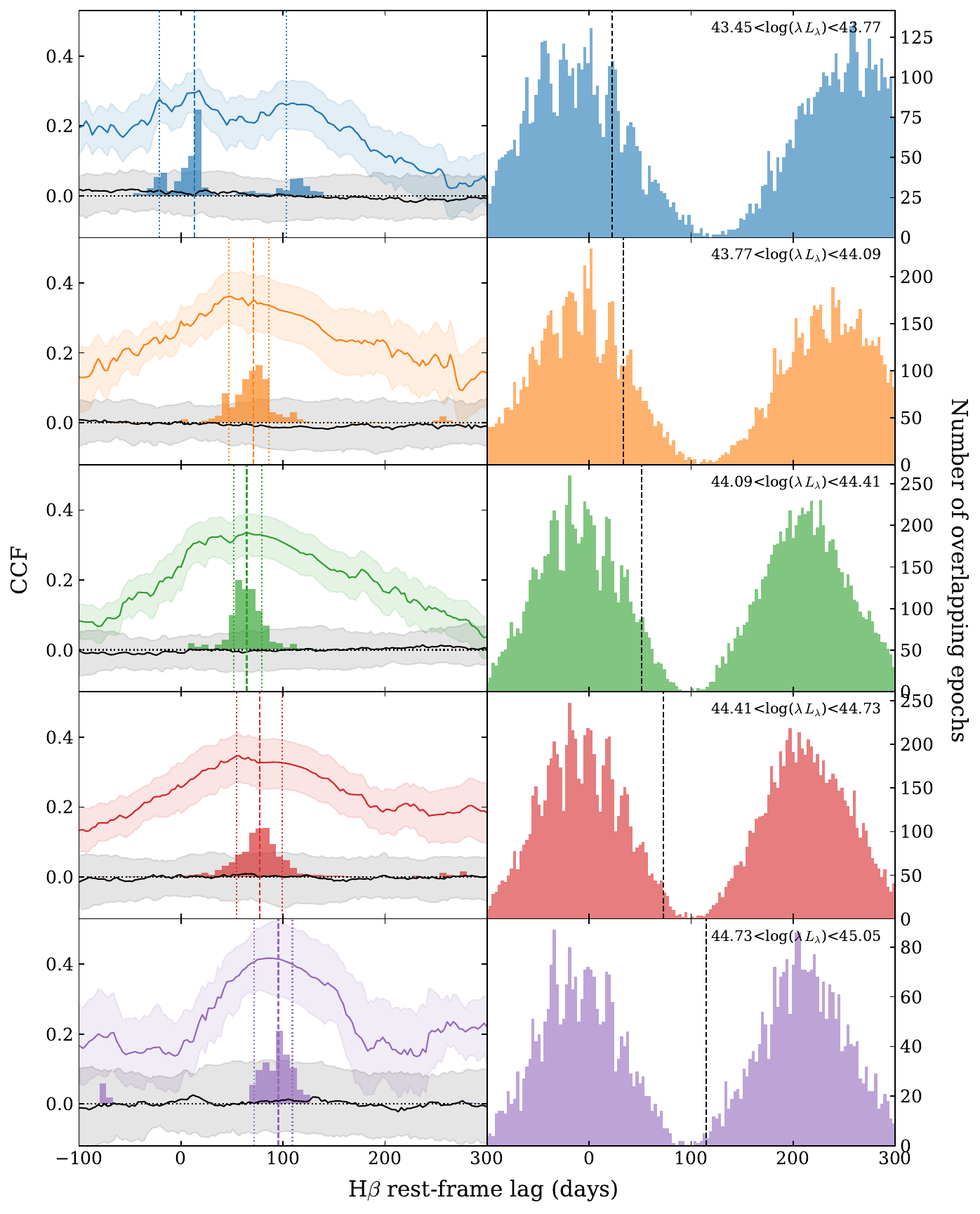}    
    \caption[]{The same as \autoref{fig:Hbeta_stackedCCF} after excluding sources with individual lag recoveries from the stacked sample. } 
    \label{fig:Hbeta_stackedCCF_r_i}
\end{figure*}

\begin{figure}
    \centering
    \includegraphics[width=\columnwidth]{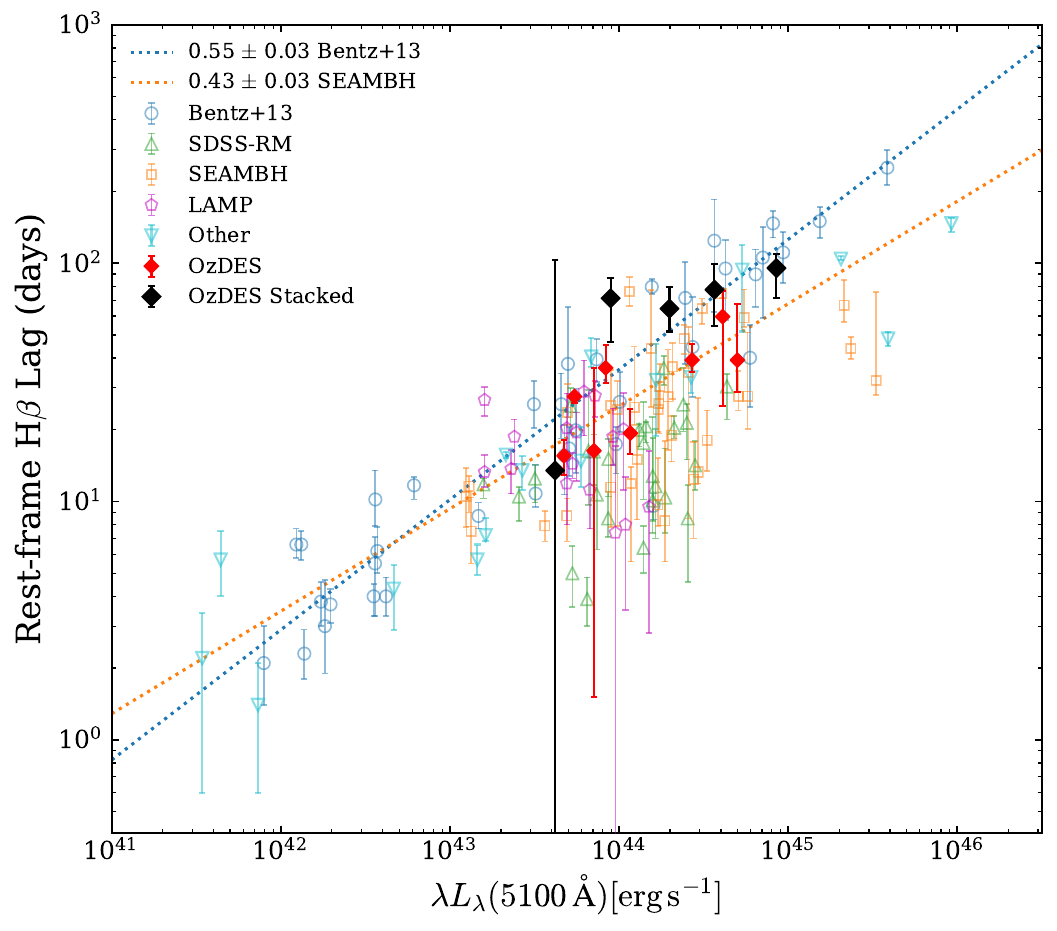} 
    \caption{The Radius-Luminosity relation for H$\beta$ (dotted lines), including the stacked average lag measurements made using the OzDES H$\beta$ sample after excluding sources with individual lag recoveries, and existing individual lag measurements (same as in \autoref{fig:Hbeta_RL}). } 
    \label{fig:Hbeta_RL_r_i}
\end{figure}

The luminosity bins for the H$\beta$ sample excluding sources with individual lags are shown in \autoref{fig:Hbeta_bins_r_i}, and the stacked CCF's and average lags for this reduced sample are given in \autoref{fig:Hbeta_stackedCCF_r_i} and \autoref{fig:Hbeta_RL_r_i}. Only the two lowest (blue and orange) and second highest (red) luminosity bins had objects excluded. The stacked CCF peaks remain intact for each of these bins, although the signal-to-noise degrades slightly, which is to expected as the size of the stacked sample decreased by $\sim$one sixth. The average lags are in close agreement to those recovered in the original analysis in \S\ref{sec:results}. This test indicates that the results from stacking are not dominated by objects with high quality individual lags, although their addition does considerably improve the lag uncertainties for the final result. 

\section*{Affiliations}
$^{1}$Research School of Astronomy and Astrophysics, Australian National University, Canberra, ACT 2611, Australia\\
$^{2}$School of Mathematics and Physics, The University of Queensland,  St Lucia, QLD 4101, Australia\\
$^{3}$Department of Astronomy, The Ohio State University, Columbus, Ohio 43210, USA\\
$^{4}$Center of Cosmology and Astro-Particle Physics, The Ohio State University, Columbus, Ohio 43210, USA\\
$^{5}$National Centre for the Public Awareness of Science, Australian National University, Canberra, ACT 2601, Australia\\
$^{6}$The Australian Research Council Centre of Excellence for All-Sky Astrophysics in 3 Dimension (ASTRO 3D), Australia\\
$^{7}$Sydney Institute for Astronomy, School of Physics, The University of Sydney, NSW 2006, Australia\\
$^{8}$Centre for Gravitational Astrophysics, College of Science, The Australian National University, ACT 2601, Australia\\
$^{9}$Laborat\'orio Interinstitucional de e-Astronomia - LIneA, Rua Gal. Jos\'e Cristino 77, Rio de Janeiro, RJ - 20921-400, Brazil\\
$^{10}$Department of Physics, University of Michigan, Ann Arbor, MI 48109, USA\\
$^{11}$Fermi National Accelerator Laboratory, P. O. Box 500, Batavia, IL 60510, USA\\
$^{12}$Departamento de F\'isica Te\'orica and Instituto de F\'isica de Part\'iculas y del Cosmos (IPARCOS-UCM), Universidad Complutense de Madrid, 28040 Madrid, Spain\\
$^{13}$Institute of Cosmology and Gravitation, University of Portsmouth, Portsmouth, PO1 3FX, UK
$^{14}$Department of Physics \& Astronomy, University College London, Gower Street, London, WC1E 6BT, UK\\
$^{15}$Instituto de Astrofisica de Canarias, E-38205 La Laguna, Tenerife, Spain\\
$^{16}$Universidad de La Laguna, Dpto. AstrofÃsica, E-38206 La Laguna, Tenerife, Spain\\
$^{17}$Institut de F\'{\i}sica d'Altes Energies (IFAE), The Barcelona Institute of Science and Technology, Campus UAB, 08193 Bellaterra (Barcelona) Spain\\
$^{18}$Centre for Extragalactic Astronomy, Durham University, South Road, Durham, DH1 3LE, UK\\
$^{19}$Hamburger Sternwarte, Universit\"{a}t Hamburg, Gojenbergsweg 112, 21029 Hamburg, Germany\\
$^{20}$Centro de Investigaciones Energ\'eticas, Medioambientales y Tecnol\'ogicas (CIEMAT), Madrid, Spain\\
$^{21}$Institute of Theoretical Astrophysics, University of Oslo. P.O. Box 1029 Blindern, NO-0315 Oslo, Norway\\
$^{22}$Kavli Institute for Cosmological Physics, University of Chicago, Chicago, IL 60637, USA\\
$^{23}$University Observatory, Faculty of Physics, Ludwig-Maximilians-Universit\"at, Scheinerstr. 1, 81679 Munich, Germany\\
$^{24}$Center for Astrophysical Surveys, National Center for Supercomputing Applications, 1205 West Clark St., Urbana, IL 61801, USA\\
$^{25}$Department of Astronomy, University of Illinois at Urbana-Champaign, 1002 W. Green Street, Urbana, IL 61801, USA\\
$^{26}$Santa Cruz Institute for Particle Physics, Santa Cruz, CA 95064, USA\\
$^{27}$Center for Astrophysics $\vert$ Harvard \& Smithsonian, 60 Garden Street, Cambridge, MA 02138, USA\\
$^{28}$Australian Astronomical Optics, Macquarie University, North Ryde, NSW 2113, Australia\\
$^{29}$Lowell Observatory, 1400 Mars Hill Rd, Flagstaff, AZ 86001, USA\\
$^{30}$George P. and Cynthia Woods Mitchell Institute for Fundamental Physics and Astronomy, and Department of Physics and Astronomy, Texas A\&M University, College Station, TX 77843, USA\\
$^{31}$Instituci\'o Catalana de Recerca i Estudis Avan\c{c}ats, E-08010 Barcelona, Spain\\
$^{32}$Observat\'orio Nacional, Rua Gal. Jos\'e Cristino 77, Rio de Janeiro, RJ - 20921-400, Brazil\\
$^{33}$Department of Physics, Carnegie Mellon University, Pittsburgh, Pennsylvania 15312, USA\\
$^{34}$Kavli Institute for Particle Astrophysics \& Cosmology, P. O. Box 2450, Stanford University, Stanford, CA 94305, USA\\
$^{35}$SLAC National Accelerator Laboratory, Menlo Park, CA 94025, USA\\
$^{36}$Department of Physics and Astronomy, Pevensey Building, University of Sussex, Brighton, BN1 9QH, UK\\
$^{37}$School of Physics and Astronomy, University of Southampton,  Southampton, SO17 1BJ, UK\\
$^{38}$Computer Science and Mathematics Division, Oak Ridge National Laboratory, Oak Ridge, TN 37831\\
$^{39}$Lawrence Berkeley National Laboratory, 1 Cyclotron Road, Berkeley, CA 94720, USA\\

\bsp	

\label{lastpage}
\end{document}